\documentstyle[12pt]{article}
 \advance\oddsidemargin by -1in
 \advance\evensidemargin
 by -1in
 \marginparsep
 0.4cm
 \advance\topmargin by
 -0.0in
 \makeindex

 \newcommand\la{\langle}
 \newcommand\ra{\rangle}
 \newcommand\beq{\begin{equation}}
 
 \newcommand\eeq{\end{equation}}
 \newcommand\beqn{\begin{eqnarray}}
 \newcommand\eeqn{\end{eqnarray}}

\begin{document}

\vspace*{1cm}

\begin{center}

{\huge\bf Energy Loss versus Shadowing\\[0.3cm] 
in the Drell-Yan Reaction
on Nuclei}

\vspace{1cm}

{\large M.B.~Johnson$^a$, B.Z.~Kopeliovich$^{b,c,d}$, 
I.K.~Potashnikova$^{b,d}$\\ and\\
P.L.~McGaughey$^a$,
J.M.~Moss$^a$, J.C.~Peng$^a$, G.~Garvey$^a$, M.~Leitch$^a$ 
C.N.~Brown$^e$,
D.M.~Kaplan$^{f,g}$
\\ \vspace*{5pt}
}

\vspace{0.5cm}

\begin{minipage}[t]{14cm}{
$^a$Los Alamos National Laboratory, Los Alamos, NM 87545, USA\\
$^b$Max-Plank-Institut f\"ur Kernphysik, 69029 Heidelberg, Germany\\
$^c$Inst. f\"ur Theor. Phys., Universit\"at Regensburg,
93040 Regensburg, Germany\\
$^d$Joint Institute for Nuclear Research, Dubna, 
141980 Moscow Region, Russia\\
$^e$Fermi National Accelerator Laboratory, Batavia, IL 60510, USA\\
$^f$Illinois Institute of Technology, Chicago, IL  60616, USA\\
$^g$Northern Illinois University, DeKalb, IL 60115, USA\\
}
\end{minipage}

\end{center}

\vspace{0.5cm}

\begin{abstract} 

We present a new analysis of the E772 and E866 experiments on the nuclear
dependence of Drell-Yan (DY) lepton pair production resulting from the
bombardment of $^2H$, Be, C, Ca, Fe, and W targets by 800 GeV/c protons
at Fermilab.  We employ a light-cone formulation of the DY reaction in the rest 
frame of the nucleus, where the dimuons detected at small values of Bjorken 
$x_2\ll 1$ may be considered to originate from the decay of a heavy photon 
radiated from an incident quark in a bremsstrahlung process.  We infer the 
energy loss of the quark by examining the
suppression of the nuclear-dependent DY ratios seen as a function of
projectile momentum fraction $x_1$ and dimuon mass $M$.  
Shadowing, which also leads to nuclear suppression of dimuons, is
calculated within the same approach employing the results of phenomenological 
fits to deep inelastic scattering data from HERA.  The analysis 
yields $-dE/dz =2.73 \pm 0.37 \pm 0.5~GeV/fm$ for the rate of quark energy
loss per unit path length, a value consistent with theoretical expectations
including the effects of the inelastic interaction of the incident
proton at the surface of the nucleus.  This is the first observation of
a nonzero energy loss effect in such experiments.

\end{abstract}

\newpage

\section{Introduction}

Nuclei can serve as a unique tool to study the space-time development of the
strong interaction during its early stages, which is inaccessible in collisions
between individual hadrons.  The Drell-Yan (DY) reaction~\cite{dy} on nuclear 
targets provides, in particular, the possibility of probing the propagation
of partons through nuclear matter in its ground state, with the produced 
lepton pair carrying away the desired information on the energy and 
transverse momentum of the parent projectile quark after it has traveled 
in the nucleus.  In this paper, we are specifically interested in the 
determination of the rate of energy loss per unit length 
$\kappa = - dE/dz$ of a fast quark 
propagating through a nucleus.  We believe that the DY reaction data provides 
the cleanest way to single out such energy-loss effects.

One might also consider using other reactions for identifying energy-loss
effects, with deep-inelastic scattering (DIS) coming to mind as one
possibility.  However, experimentally \cite{osborne,emc,e665,hermes}, it has 
proven difficult to identify the partonic energy loss, not to mention to
specify how to best measure this potentially important phenomenon.  The origin 
of the difficulty can be appreciated by considering the following experiment.  
Suppose one were to measure the nuclear (A) dependence of semi-inclusive DIS, 
with known 4-momentum transfer.  In this experiment, the A dependence of the 
energy (momentum) carried by a hadron created by the struck quark is supposed
to convey the information regarding parton energy loss.  However, besides the 
vacuum (initiated by the DIS) and induced (due to multiple interactions of the 
struck parton) energy loss, one must also account for the hadronization 
parton $\to$ hadron.  As a rule, this hadron is subject to absorption when 
it is produced inside the nucleus.  The complicated dynamics of this 
(see \cite{knp} and the comparisons of the predictions to data \cite{hermes}) 
makes it extremely difficult to single out the net effect of energy loss.

High-$p_T$ production of hadrons off nuclei is even more complicated, since it 
includes a convolution of the parton distribution in the incident hadron
with the high-$p_T$ parton-parton scattering cross section.  Multiple parton 
interactions inside the nuclear medium (Cronin effect) also makes the 
interpretation extremely complicated.  Yet another reaction sensitive to 
energy loss is charmonium production off nuclei. The violation of $x_2$ 
scaling observed in the E772  experiment \cite{e772-psi} has already suggested 
the presence of energy-loss effects. However, final state absorption and 
coherence effects (see \cite{kth}) do not allow a clear identification of 
energy loss in the data.

In the DY reaction, the ratios $R_{A/D}$ of the cross section for a heavy 
nucleus $A$ compared to a light one (the deuteron $D$, say), are particularly 
relevant for the study of parton energy loss in nuclei.  This is especially 
true in the region of large values of the the longitudinal momentum fraction 
of the projectile hadron carried by the DY pair, where significant nuclear 
suppression is clearly seen in this ratio.  In an early study~\cite{kn1}, it 
was shown that a ratio such as $R_{A/D}$ would be sensitive to $\kappa$ 
because the fraction $x_q$ of the light-cone momentum of the incoming hadron 
$h$ carried by projectile quark $q$ shifts to smaller values in $hA$ 
collisions as a result of initial-state energy loss.  In the case of the DY 
reaction, this shift $\Delta x_q$ suppresses the ratio $R_{A/D}$ because it 
results in the sampling of the parton distribution at larger $x_q$ (for the 
same momentum of the dilepton), where the projectile parton densities are 
smaller.  Thus, the ratio $R_{A/D}$ must decrease towards large Feynman 
$x_F\to 1$ where energy loss leads to a strongest suppression. 

At the same time, nuclear shadowing was observed in Ref.~\cite{kn1} to be 
a competing source of suppression at large $x_F$ (more specifically, small 
values of the momentum fraction $x_2$ of the target).  As it is easy to mix up 
energy loss and shadowing, one must take special precautions to disentangle 
these two effects when analyzing experimental data.

The first Drell-Yan data suitable for such an analysis was obtained in 
Fermilab E772/E866 experiments.  An analysis of the E772 data~\cite{e772}
was made in \cite{gm} ignoring shadowing and assuming that $\kappa$ rises 
linearly with energy. The latter assumption  was criticized~\cite{bh} for 
violating the Landau-Pomeranchuk principle.

A better analysis was performed recently by the the E866 collaboration using 
the E866 data~\cite{vasiliev}.  The analysis of Vasiliev, {\it et al.},
which attempted to improve on that of \cite{gm} by including shadowing, 
considered three
scenarios for energy loss, one of which was the same as in \cite{gm}.
All three of these gave $\kappa$ consistent with essentially zero energy 
loss, in contrast to the value $\kappa \approx 1.5 Gev/fm$ found in \cite{gm}.
Thus, shadowing was found in \cite{vasiliev} to be the main source of 
nuclear suppression
of the DY cross section at large $x_F$.

Our present work differs from the previous analyses in that we attempt to 
unambiguously separate shadowing and energy loss by calculating the shadowing 
correction to the DY data using theory.  The concept of coherence length, 
which plays a key role in respecting the Landau-Pomeranchuk principle, is 
essential to this formulation. It is our belief that the procedure 
employed by Vasiliev, {\it et al.} overestimated the shadowing contribution 
and hence substantially underestimated $\kappa$.  The reason is that the 
shadowing correction was taken from a phenomenological analysis~\cite{eks} 
that had already attributed the suppression of $R_{A/D}$ observed in E772 data 
at large $x_F$ entirely to shadowing.  Preliminary 
results of our analysis are reported in Ref.~\cite{johnson}.

We describe the interplay between quark energy loss and shadowing by working 
in the target rest frame using the light-cone approach of 
Refs.~\cite{hir,bhq,kst1,krt3}, where these are given as separate 
contributions to the DY cross section.  The target rest frame formulation 
is discussed in Sect.~2.1, where we 
relate the DY process to projectile fluctuations containing the DY pair.
In terms of such fluctuations, the DY reaction may be viewed as occurring when 
interactions with the target break the coherence of the fluctuation and free a 
DY pair, {\it i.e.} bringing it on its mass shell. Such an interpretation is 
quite different from the conventional partonic treatment of DY process as 
annihilation $\bar qq \to \bar ll$. However, the partonic interpretation is 
known to be Lorentz noninvariant and vary from frame to frame.
  
Such a rest frame formulation bears a close analogy to the more familiar
light-cone description of deep-inelastic scattering (DIS) at small $x$ 
\cite{nz91}. In that case, the incident virtual photon develops 
a quark-antiquark fluctuation which is brought to its mass shell
by the interaction with the target. Again, this picture looks very 
different from the conventional parton model interpretation in the Breit
frame, where it is described as absorption of the incident virtual photon by 
a quark (antiquark), which in this case belongs to the target. 

The two processes, DY and DIS are known to be closely related in the parton
model via QCD factorization. It is not surprising that the rest frame 
descriptions of these reactions also look similar. One of the advantages 
of this approach is the clear and simple treatment of nuclear shadowing,
which is described in terms of the usual Glauber-Gribov theory. Shadowing
in both processes, DY and DIS, is controlled by the same universal color-dipole
cross section \cite{zkl}, which is the cross section for freeing the 
fluctuations. The strategy used in present paper is to treat the dipole cross 
section phenomenologically by fitting it to DIS data and then verifying the 
theory by comparing it to DIS scattering data on nuclei in the shadowing 
region (where there is no danger to mix up shadowing with energy loss).
Then one can safely predict shadowing for the DY process. Such a strategy
is based on the universality of the dipole cross section, which is
a manifestation of QCD factorization.
 
The lifetime of the 
fluctuations, or coherence length, is another 
crucial quantity for understanding how 
energy loss and shadowing occur in the target rest frame.  The coherence 
length is discussed and calculated in Sect.~\ref{lc}.  This turns out to be 
mainly a function of Bjorken $x_2$, with some corrections that dramatically 
deviate from QCD factorization towards the smallest $x_2$.  

We discuss the use of the DY reaction as a probe for quark energy loss in 
Sect.~\ref{eloss}.  Section \ref{models} gives an overview of model 
expectations for energy loss.  
The string model (Sect.~\ref{string1}) predicts a constant rate of energy
loss following the first inelastic collision of the incident hadron on the
nuclear surface.  The magnitude of this is about the same as that of the
string tension, $\kappa_s\approx 1\,GeV/fm$.  Multiple interactions of the
quark in the nuclear medium lead to an additional induced energy loss whose
rate rises linearly with the path length of the quark.  
Numerically
this is a small correction to the dominant constant term.  Similar effects
follow from perturbative QCD (Sect.~\ref{pqcd}).  The first inelastic
interaction of the incident proton in the nuclear surface initiates a
long-lasting gluon bremsstrahlung, providing a constant rate of energy loss
of about the same value as that given by the string model.  Energy loss induced
by quark rescattering also rises quadratically with the length of the path
and is of similar value as in the string model.  These two sources of energy
loss are complementary and must be added, giving an expected rate of energy
loss of about $2~GeV/fm$.

In the target rest frame, the suppression of $R_{A/D}$ by $\kappa$ is 
associated 
with short-lived fluctuations in which the DY pair is released immediately
after the interaction of the projectile quark with a target nucleon.
The produced dilepton pair carries undisturbed information
about the energy of the quark traversing the nucleus only for such
fluctuations.
An important, although simple, consideration that arises here is the 
determination of the length of the path in nuclear matter
over which the parton in the initial state loses energy.  It is demonstrated 
in Sect.~\ref{path} that the mean length of this path does not follow an
$A^{1/3}$ dependence but is, in fact, quite short compared to the nuclear 
radius.  The distribution over path length is calculated for different nuclei.
Nuclear suppression caused by energy loss is then given as a convolution of the 
quark distribution function of the incident hadron, modified by energy loss,
with the DY cross section 
for the quark-nucleon interaction (Sect.~\ref{ratio}).  The former is 
borrowed from phenomenological models successfully describing soft hadronic 
collisions (Sect.~\ref{f_q}), while the latter is fitted to data from the 
E772 experiment on a deuterium target (Sect.~\ref{qN}).  

The other consideration in understanding the suppression of $R_{A/D}$ is, as 
we have remarked, the shadowing process.  The critical quantities 
for describing shadowing in this approach 
are the coherence length and the effective fluctuation-freeing cross section 
(averaged over different fluctuations).  The effective cross section is 
calculated in Sect.~\ref{shadowing} in terms of a color-dipole cross 
section that describes
data from HERA for the proton structure function at high $Q^2$. 
The shadowing correction is then calculated in an eikonal description of the
multiple scattering of the incident quark in terms of the color-dipole
cross section.  In 
Sect.~\ref{shadowing} we also confirm explicitly our description of shadowing 
by comparing the theory to deep inelastic scattering data, where energy loss 
is not an issue.  We make our determination of $\kappa$ 
from $R_{A/D}^{expt}(x_1,M^2)$ based on the theoretical description of the 
effects of shadowing in Sect.~\ref{eloss-shad}.  We fit the parameter $\kappa$ 
to the entire set experimental data without preselection, and the 
result is in agreement with the expected value of $\kappa$.

We also performed a variety of tests to check the stability of our results.
These tests are described in Sect.~5.2.  In one, we examined the sensitivity 
of the rate of energy loss to the relative contribution of shadowed and 
nonshadowed events.  We determined that selecting only those events with small 
$x_2$, where shadowing is the dominant effect, does not change the results of 
the fit.  We also artificially enlarged or eliminated shadowing, corrected
shadowing calculations for multiple interactions, etc., and found stable
results.  These tests also provide a scale for the systematic uncertainties in 
our analysis.

Section~5.3 is devoted to the important issue of the disagreement between our
determination of the rate of energy loss and that of a previous analysis of 
the E866 data for the DY reaction~\cite{vasiliev}, which detected no 
energy-loss effect.  In Sect.~\ref{trf} we discuss why the choice of the 
target rest frame is the natural one for formulating the theory of energy loss.
In Sect.~5.3b we find that the different conclusions reached in 
Ref.~\cite{vasiliev} can be explained 
by the substantially different space-time variation in the DY reaction at 
small $x_2$ and the larger density of antiquarks at large $x_2$ employed in 
this work.  We justify our own results based on detailed theoretical arguments.

In Sect.~6 we summarize the results of the present analysis.

\section{Drell-Yan reaction in the nuclear rest frame}\label{dy-sect}

\subsection{\boldmath$\bar qq\to \bar ll$ annihilation or 
\boldmath$q\to q\bar ll$ bremsstrahlung?}

\begin{figure}[tbh]
\includegraphics{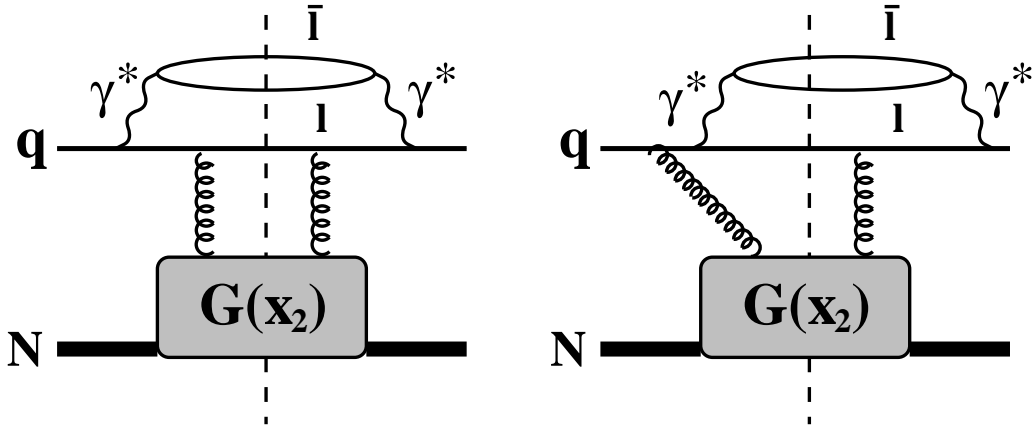}
\begin{center}
\vspace{6cm}
\parbox{13cm}
{\caption[Delta] {\sl
Feynman diagrams for the cross section of lepton pair production in 
a quark-nucleon collision with the $\gamma^*$ radiated before (left)
and after (right) the interaction with the target.}
\label{dy}}
\end{center}
\end{figure}

In the target rest frame, the Drell-Yan reaction at small $x_2$ corresponds to 
the electromagnetic radiation of a lepton pair by a projectile quark (valence 
or sea, depending on the value of $x_1$), rather than to $q\bar q\to l\bar l$ 
annihilation \cite{hir}.  In this case, the $l\bar l$ pair is imagined to be 
liberated from a virtual fluctuation of the projectile when it interacts with 
a nucleon of the nucleus. Two examples of Feynman diagrams contributing to 
this are shown in Fig.~\ref{dy}. 
The cross section corresponding to these and other diagrams 
not shown has a factorized form in the impact parameter representation 
\cite{hir,bhq,kst1},
\beq
\frac{d\sigma^{qN}_{DY}}
{dM^2\,d\alpha} = \int
d^2r_T\,\Bigl|\Psi_{q\bar ll}(r_T,\alpha)\Bigr|^2\,
\sigma_{\bar qq}(\alpha r_T,s_{qN})\ ,
\label{0.1}
\eeq
where $s_{qN}$ is the square of the center-of-mass energy of the quark-nucleon 
collision, $M$ is the dilepton effective mass, 
and $\Psi_{q\bar ll}(r_T,\alpha)$ is the light-cone wave function 
of the $|q\bar ll\ra$ Fock component of the projectile quark.  This 
wave function depends on both the transverse separation $r_T$ between the 
$\bar ll$ and the recoil quark and the fraction $\alpha$ of the initial 
light-cone momentum carried by the $\bar ll$ pair. Here,
$\sigma_{\bar qq}(\rho,s)$ is the universal dipole cross section
for the interaction of a colorless dipole $\bar qq$ with transverse separation
$\rho=\alpha r_T$ from the nucleon, introduced in \cite{zkl}. The appearance
of the color dipole cross section in Eq.~(\ref{0.1}) arises from the
difference between the interaction for $|q\ra$ and $|q\bar ll\ra$
fluctuations.  This means, among other things, that the DY cross section 
receives no contribution if the transverse positions of the initial and 
final quarks coincide, and at small $\rho$ this cross section vanishes 
$\propto \rho^2$ \cite{zkl}. Said otherwise, when $r_T \to 0$, the strong 
interaction cannot discriminate between the Fock components $|q\ra$ and 
$|q\bar ll\ra$ and therefore radiation is not possible.

As usual for
bremsstrahlung, one cannot say whether the (virtual) photon is radiated
before or after the interaction with the target (via gluon exchange); 
both are represented in Fig.~\ref{dy}. This quantum-mechanical 
uncertainty in the time of radiation is called the coherence time, which is
the same as the coherence length since we are near the light cone. 
In terms of the light-cone approach, the coherence time can also be 
interpreted as the lifetime of the 
$|q\bar ll\ra$ fluctuation of the incoming quark.

As already noted, one may distinguish this approach from the usual partonic 
interpretation of 
the DY reaction noting that the space-time development of high-energy reactions 
treated in terms of the parton model is not Lorentz invariant and depends
on the reference frame.  Indeed, since 
even the time ordering varies from frame to frame, the lepton pair produced at 
small values of Bjorken $x_2\ll 1$ in the DY reaction may be equivalently 
viewed as annihilation $\bar qq \to \bar ll$ in the rest frame of the 
photon~\cite{dy} or as the decay of a heavy photon radiated in a 
bremsstrahlung process.

\subsection{Coherence length}\label{lc}

The coherence length is an important quantity controlling  nuclear effects in 
the DY reaction.  If the coherence length is longer than the mean internucleon 
separation in the nucleus, different bound nucleons compete in freeing the 
$\bar ll$ fluctuation.  This phenomenon is known as nuclear shadowing.  On the 
other hand, if the coherence length is very short, the fluctuation has time to 
interact with only one bound nucleon. In this case, all nucleons contribute 
equally to the DY cross section, {\it i.e.} there is no shadowing.

The coherence length for a fluctuation of a projectile quark $q\to \bar ll\,q$
is given by the energy denominator,
\beq
l^{DY}_c=\frac{p_q^+}{M^2_{\bar llq}-m_q^2}\ ,
\label{1}
\eeq
where
\beq
M^2_{\bar llq}=\frac{M^2}{1-\alpha}+\frac{m_q^2}{\alpha}
+ \frac{k_T^2}{\alpha\,(1-\alpha)}\ 
\label{2}
\eeq
is the effective mass squared of the $\bar llq$ fluctuation;
$p_q^+=E_q+p^{||}_q$ is the light-cone momentum of the incident quark; 
$M$ is the dilepton effective mass; and, $\vec k_T$ and $\alpha$
are the transverse momentum and fraction of the light-cone
momentum of the parent quark carried by the $\bar ll$ pair, respectively.

If energy is conserved, the longitudinal momentum transfer between the
initial state $|q\ra$ and the fluctuation $|q\bar ll\ra$ is $q_c=1/l_c$.  
Thus, one can say that the coherence length is the maximal longitudinal
distance between fluctuations that are in phase.

The DY variables $x_1$ and $x_2$ satisfy the equations,
\beq
x_1\,x_2=\frac{M^2}{s},
\label{2a}
\eeq
\beq
x_1-x_2=x_F\ ,
\label{2b}
\eeq
where $x_F$ is the Feynman variable, and they are interpreted in the parton
model as the Bjorken variables of the annihilating quark and antiquark.
They can equivalently be defined as fractions of the light-cone
momenta $P^{\pm}=E\pm P^{||}$ of the beam (b) and a target (t) nucleon
carried by the lepton pair as follows,
\beq
x_1=\frac{p^+_{\bar ll}}{P_b^+}\ ,\ \ \ \ \ \
x_2=\frac{p^-_{\bar ll}}{P_t^-}\ .
\label{2c}
\eeq
Therefore $p^+_{\bar ll} = \alpha\,p_q^+=x_1\,P_b^+$, and the coherence
length (\ref{1}) 
averaged over $\alpha$ and $k_T$ reads
\beq
\left\la l^{DY}_c\right\ra
=\frac{\left\la K^{DY}\right\ra_{M^2,x_1}}{m_N\,x_2}\ ,
\label{3}
\eeq
where $\la K^{DY}\ra_{M^2,x_1}$ is
\beq
\left\la K^{DY}\right\ra_{M^2,x_1}\equiv
\left\la K^{DY}(\alpha,k_T)\right\ra_{\alpha,k_T}
=\left\la\frac{M^2\,(1-\alpha)}
{M^2\,(1-\alpha)+\alpha^2\,m_q^2+k_T^2}
\right\ra_{\alpha,k_T}\ .
\label{4}
\eeq
In perturbative calculations, one would take $m_q$ to be the current 
quark mass.  However, this is unsatisfactory in the present case because it 
leads to large transverse separations between the parent and recoil quark when 
the radiated dilepton takes essentially the entire initial quark momentum 
$M^2(1-\alpha)\sim m_q^2$.  Indeed, in this case the separation becomes
$r_T \sim 1/m_q$, and one can grossly overestimate the contribution of large 
separations when using the approximation $\sigma_{\bar qq}(r_T)\propto r_T^2$.  
This divergence can be regularized by introducing an effective quark mass that 
suppresses the probability of large separations.  Experience with 
DIS~\cite{krt2} shows that employing $m_q \approx 0.2\,GeV$ gives a good 
description of DIS data.  We use the same effective mass in what follows for 
the DY reaction.  Of course, this divergence does not create any problem when 
the dipole cross section saturates at large separations.  

Clearly the coherence length, which controls nuclear shadowing, may not 
scale in $x_2$ and differs from the usually-used approximation 
$l_c=1/2x_2m_N$.  Moreover, it follows from (\ref{4}) that as $x_1\to 1$, 
$l_c\to 0$ (since $\alpha > x_1$), {\it i.e.} nuclear shadowing vanishes. 
This fact is at variance with the usual expectation that shadowing 
increases towards the limit $x_1=1$, which corresponds to the smallest $x_2$.
Although this contradiction shows that QCD factorization breaks down 
dramatically at large $x_1$, the result is quite consistent with the 
observation that all physics becomes soft~\cite{bmht} at large $x_1$.

The averaging in (\ref{4}) should be weighted by the light-cone wave function
of the $q\bar ll$ fluctuation squared, which is known to diverge (for
transversely polarized virtual photons) at small transverse $q-\bar ll$
separation $r_T$, {\it i.e.} for large $k_T$ \cite{kst1} (similar to the case
of deep inelastic scattering (DIS) for the fluctuation $\gamma^*\to\bar qq$).  
Such fluctuations with small $r_T$ do not interact; therefore, to avoid the 
nonsensical result that $\la K^{DY}\ra\equiv 0$, which would arise from the 
fact that the vacuum fluctuations are dominated by large $k_T$ -- leading to a 
divergent normalization -- one also needs to include $\sigma_{\bar qq}(r_T,s)$
\cite{hir,kst1} in the evaluation of Eq.~(\ref{4}).  It is also necessary to 
include it here since we are 
interested only in those fluctuations that participate in the interaction.  
The dipole cross section plays the same role in Eq.~(\ref{0.1}), namely it 
prevents the interaction
from discriminating between the fluctuation $|q\ra$ and $|q\bar ll\ra$ as
$r_T \to 0$.  A distinction between the two Fock components is clearly needed 
in order to liberate the $\bar ll$ pair.

In what follows, we are interested in nuclear shadowing, which in lowest 
order originates from double scattering.  For reasons given above, the 
relevant mean coherence length should be weighted with 
$\sigma^2_{\bar qq}(\alpha r_T,s)$.  Thus, Eq.~(\ref{4}) can be written 
explicitly as,
\beq
\left\la K^{DY}\right\ra_{M^2,x_1} = 
\frac{\int\limits_{x_1}^{1}
d\,x_q\,F^h_q(x_q)\,
\int d^2 k_T\,
\left|\widetilde\Psi_{\bar llq}
(x_1/x_q,k_T)\right|^2\,
K^{DY}(x_1/x_q,k_T)}
{\int\limits_{x_1}^{1}
d\,x_q\,F^h_q(x_q)\,
\int d^2 k_T\,
\left|\widetilde\Psi_{\bar llq}
(x_1/x_q,k_T)\right|^2}\ ,
\label{5}
\eeq
where $F^h_q(x_q)$ is the quark distribution function of the beam hadron,
which depends on the fraction $x_q$ of the hadron light-cone momentum carried 
by 
the quark.  Hereafter we restrict ourselves to the transversely polarized 
$\bar ll$ pairs dominating the DY cross section, noting that the contribution 
of longitudinal polarization vanishes in any case as $x_1\to 1$. The modified 
light-cone distribution amplitude $\widetilde\Psi^T_{\bar llq}$ 
reads~\cite{krt2}
\beq
\widetilde\Psi_{\bar llq}^{T}
(\alpha,k_T)=
\int d^2r_T\,e^{i\,\vec r_T\cdot\vec k_T}\,
\sigma_{\bar qq}(\alpha r_T,s)\,
\Psi_{\bar llq}^{T}(\alpha,r_T) ,
\label{6}
\eeq
where \cite{hir,bhq,kst1}
\beq
\Psi^{T}_{\bar llq}(\alpha,k_T)=
Z_q\,\frac{\sqrt{\alpha_{em}}}{2\,\pi}\,
\chi_f\,\widehat O^{T}\,\chi_i\,K_0(\tau r_T) .
\label{7}
\eeq
Here $\chi_{i,f}$ are the spinors of the initial and final quarks, $Z_q$ is
their charge,
$K_0(x)$ is a modified Bessel function, and
\beq
\tau^2 =
(1-\alpha)\,M^2
+\alpha^2\,m_q^2\ .
\label{8}
\eeq
The operator $\widehat O^{T}$ has the form 
\beq
\widehat O^{T} = i\,m_q\alpha^2\,
\vec {e^*}\cdot (\vec n\times\vec\sigma)\,
 + \alpha\,\vec {e^*}\cdot (\vec\sigma\times\vec\nabla)
-i(2-\alpha)\,\vec {e^*}\cdot \vec\nabla\ ,
\label{9}
\eeq
where $\vec e$ is the polarization vector of the $\bar ll$ (virtual photon),
$\vec n$ is the unit vector along the projectile momentum of the quark,
and $\vec\nabla$ acts on $\vec r_T$.

\begin{figure}[tbh]
\includegraphics{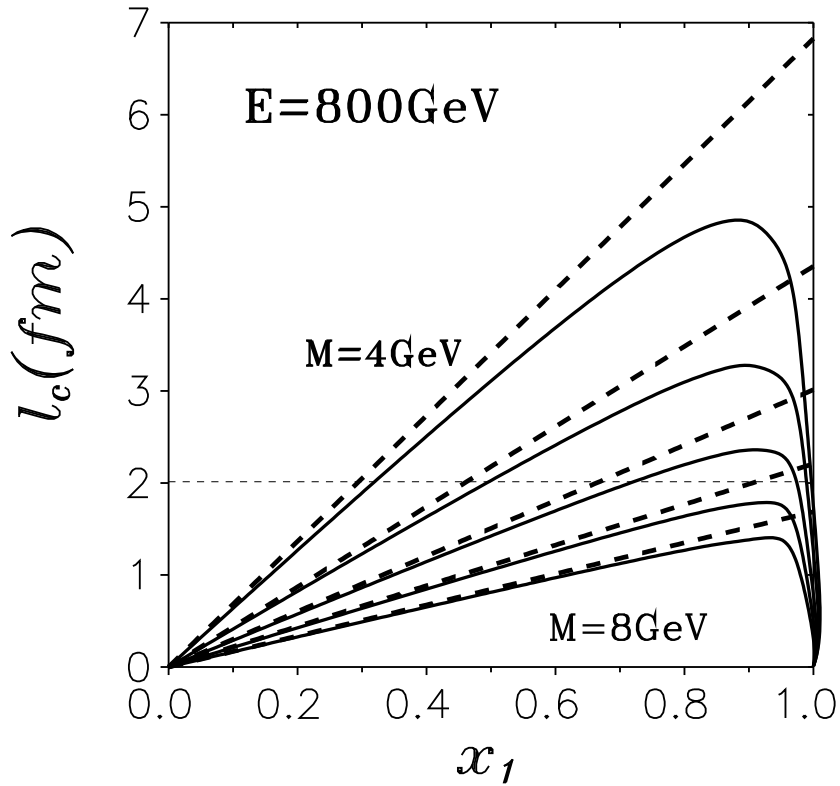}
\begin{center}
\vspace{12cm}
\parbox{13cm}
{\caption[shad1]
{\sl The mean coherence length (\ref{3}) as function of $x_1$
and dimuon effective mass $M=4,\ 5,\ ...\ 8\ GeV$
(solid curves). Dashed lines show the predictions of QCD 
factorization which relates the cross section of the DY reaction
and DIS on a nucleus.}
\label{lc-r2}}
\end{center}
\end{figure}

To simplify calculations we use the dipole cross section in the form 
corresponding to small $r_T$ (a more realistic shape, leveling off at large 
$r_T$~\cite{gw,kst2}, leads to similar results), 
\beq
\sigma_{\bar qq}(r_T,s) = C(s)\,r_T^2 ,
\label{10}
\eeq
where the constant of proportionality $C(s)$ does not enter into the final 
results. 
Then, the modified distribution amplitude in $k_T$-representation 
has the form,
\beq
\widetilde\Psi^T_{\bar llq}(\alpha,k_T)=
2\,Z_q\,\sqrt{\alpha_{em}}\,C(s)\,\vec e\cdot\vec k_T\,
\frac{i\,\alpha^2\,\tau^2}{\pi\,(\tau^2+k_T^2)^3}\ ,
\label{11}
\eeq
and the integration over $k_T$ in (\ref{5}) can be
performed analytically,
\beq
\left\la K^{DY}\right\ra=\frac{2\,M^2}{3}\,\,
\frac{N^{DY}}{D^{DY}}\ ,
\label{11a}
\eeq
where
\beq
N^{DY}=
\int\limits_{x_1}^{1}
d\,\alpha\,\,
F^h_q(x_1/\alpha)\,\frac{
(1-\alpha)\,\alpha^2\,[1+(1-\alpha)^2]}
{\tau^6}\ ;
\label{12}
\eeq
\beq
D^{DY}=
\int\limits_{x_1}^{1}
d\,\alpha\,\,
F^h_q(x_1/\alpha)\,
\frac{\alpha^2\,
[1+(1-\alpha)^2]}
{\tau^4}\ .
\label{12a}
 \eeq
 The coherence length $l_c^{DY}$ calculated with (\ref{3}) and (\ref{11a}) -
(\ref{12a}) is plotted by solid curves as function of $x_1$ in
Fig.~\ref{lc-r2} for different dilepton masses $M=4,\ 5,\ ...\ 8\ GeV$.

Equation~(\ref{11a}) should be compared with the analogous factor 
$\la K^{DIS}\ra = m_Nx_2\la l_c^{DIS}\ra$ for DIS
calculated in Ref.~\cite{krt2},
\beq
\left\la K^{DIS}\right\ra = 
\frac{2\,Q^2}{3}\,\,
\frac{N^{DIS}}{D^{DIS}}\ ,
\label{12b}
\eeq
where
\beq
N^{DIS} =
\int\limits_{0}^{1}
d\,\alpha\,\,
\frac{\alpha(1-\alpha)
[\alpha^2+(1-\alpha)^2]}
{\epsilon^6}
\label{13}
\eeq
\beq
D^{DIS} =
\int\limits_{0}^{1}
d\,\alpha\,\,
\frac{\alpha^2+(1-\alpha)^2}
{\epsilon^4}\ ,
\label{13a}
\eeq
and where
\beq
\epsilon^2=\alpha\,(1-\alpha)\,Q^2 
+ m_q^2\ .
\label{14}
\eeq
Obviously, $\la K^{DIS}\ra$ depends only on $Q^2$, while $\la K^{DY}\ra $ 
depends on $M^2$ and $x_1$ and, as we have noted, vanishes as $x_1\to 1$. 

We compare in Fig.~\ref{lc-r2} the coherence length $\la l_c\ra =\la
K\ra/m_Nx_2$ calculated with (\ref{11a}) and (\ref{12b}) for kinematics
corresponding to the E772/E866 experiments.  The dashed curves are
calculated for DIS at the same $x_{Bj}=x_2$.
The discrepancy between the two sets of curves, which increases
towards $x_1=1$, manifests a deviation from factorization.  Nevertheless, 
factorization  is restored at small $x_1$ and/or at large $M$. Note that the 
factorization theorem requires only that the soft physics, which is 
common to DIS and DY, should factor from the reaction mechanism at large
$Q^2$. Since shadowing in the distribution function is controlled in
the target rest frame formulation by the coherence length and an effective
cross section ($\sigma_{eff}$ discussed in Sect.~\ref{shadowing} below),
factorization requires that these two quantities should also be the same
for DIS and DY in this limit.

\begin{figure}[tbh]
\includegraphics{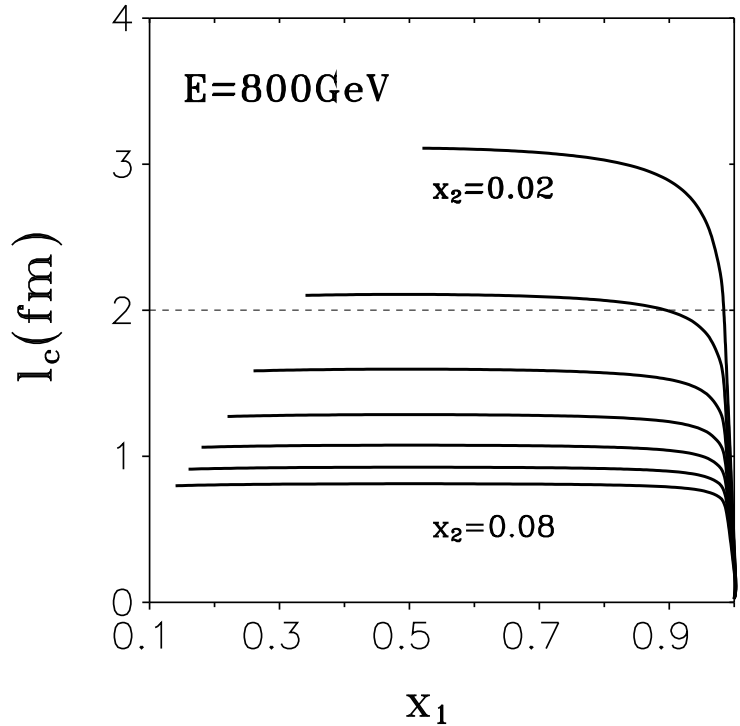}
\begin{center}
\vspace{9cm}
\parbox{13cm}
{\caption[shad1]
{\sl The mean coherence length (\ref{3}) as function of $x_1$
at fixed values of $x_2=0.02,\ 0.03,\ ...\ 0.08$.}
\label{lc-x2}}
\end{center}
 \end{figure}

Note that the $x_2$ dependence of the mean coherence length is controlled 
mainly by the denominator in (\ref{3}), while the factor 
$\la K^{DY}\ra$ is a rather flat function of $x_2$ and $x_1$. 
This is demonstrated in Fig.~\ref{lc-x2} where $\la l_c^{DY}\ra$ is plotted 
versus $x_1$ for different fixed values of $x_2$.  
 \begin{figure}[tbh]
\includegraphics{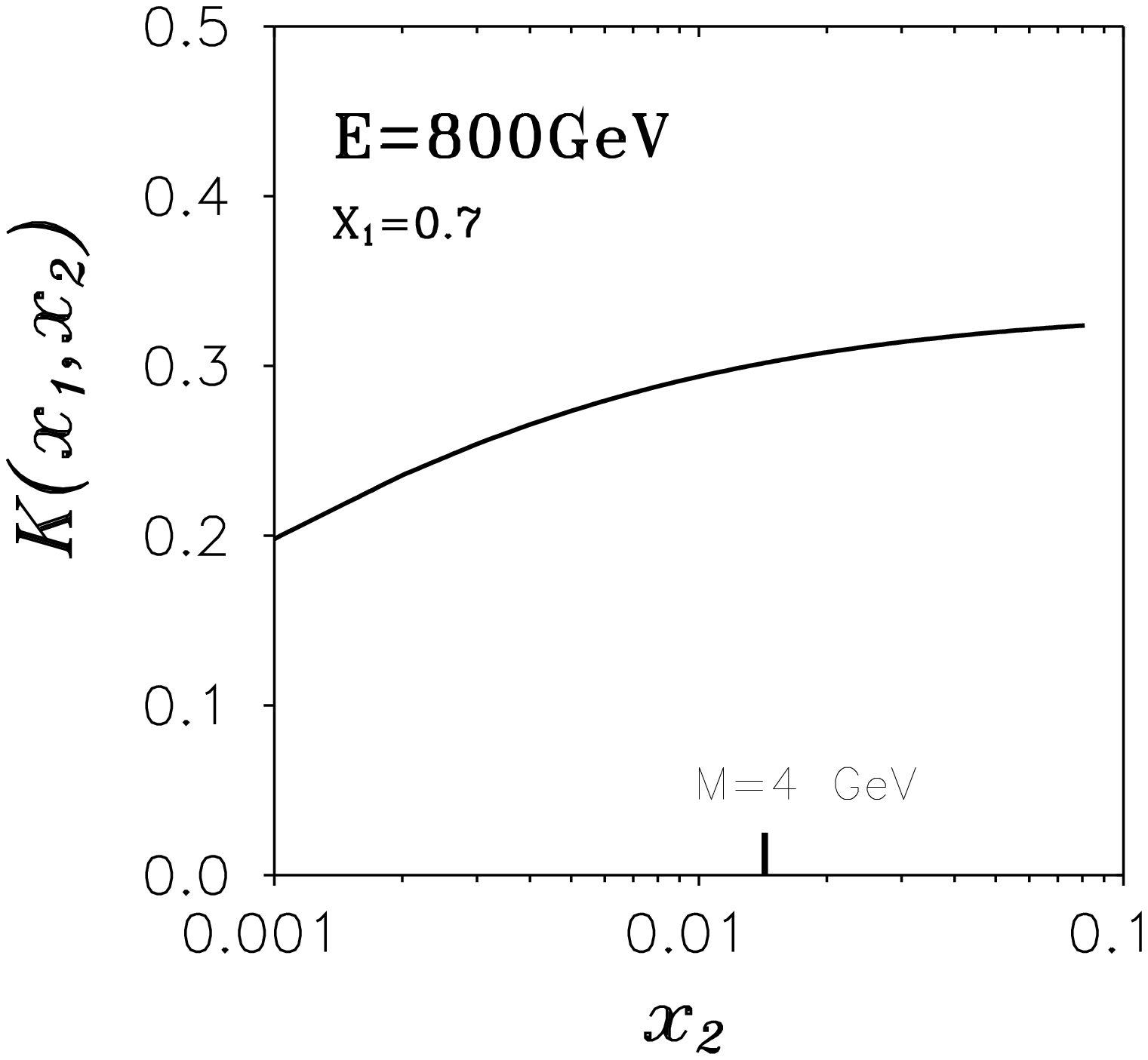}
\begin{center}
\vspace{9cm}
\parbox{13cm}
{\caption[shad1]
{\sl Factor $K^{DY}$ appearing in the expression 
for $\left\la l^{DY}_c\right\ra$ (Eq.~(\ref{3})) as function of $x_2$
at $x_1=0.7$.}
\label{k-x2}}
\end{center}
\end{figure}

At fixed $x_2$ the coherence length is nearly constant within the kinematical
range of the E772/E866 experiments at $x_1 < 0.8$.  The $x_2$ dependence of
$\la K^{DY}\ra_{x_1,x_2}$ at $x_1=0.7$, depicted in Fig.~\ref{k-x2},
demonstrates that $\la K^{DY}\ra = 0.2 - 0.3$ (see also in \cite{krt2}) is
quite small compared to the usually accepted $\la K^{DY}\ra =0.5$. In the
kinematical range of the E772/E866 experiments, $M > 4\,GeV$, and $\la
K^{DY}\ra\approx 0.3$ does not vary much.

\section{Energy loss of the projectile quark in nuclear matter}
\label{eloss}

The coherence among the soft partons of the incident proton is destroyed as a 
result of inelastic interactions of the projectile in the surface of the 
nucleus.  These partons then propagate through the nucleus independently, 
losing energy (to hadronization) and having their transverse momentum 
broadened by interactions.  This is predominantly a nonperturbative process, 
since the projectile quark contains mainly soft fluctuations.  

Eventually, one of these quarks may develop a hard fluctuation containing the 
lepton pair such that the lifetime of the fluctuation is short compared to the 
mean spacing between bound nucleons.  The $\bar ll$ pair is immediately produced
when a hard interaction occurs with a bound nucleon within the very short 
fluctuation time that we have referred to as the coherence time.  Any change 
in the longitudinal or transverse momentum of the projectile quark caused by 
the preceding initial state interactions can diminish the energy of the 
projectile quark participating in the DY reaction.  This affects the momentum
spectrum for the produced DY pair, which then serves as a probe for the dynamics
of the initial state interaction.  It corresponds to the first of the two 
sources of nuclear suppression of $R_{A/D}(x_F)$ mentioned in the introduction. 

\subsection{Models for energy loss}\label{models}

\subsubsection{String model}\label{string1}

The first inelastic interaction of the beam in the surface of the nucleus
via color exchange with a bound nucleon leads to the formation of color strings
between the target and beam partons. Due to the constant retarding action
of the string, the leading projectile quark loses energy with a constant
rate per unit length, $dE/dz=-\kappa_s$ \cite{kn1,kn2}, which is invariant
relative to longitudinal Lorentz boosts. The string tension $\kappa_s$ is
related to the slope parameter $\alpha_R^{\prime}$ of the mesonic 
Regge trajectories~\cite{cnn} as
 \beq
\kappa_s=\frac{1}{2\,\pi\,\alpha_R^{\prime}}
\approx 1\,GeV/fm\ .
\label{3.0}
\eeq
This value imposes a scale for the expected rate of energy loss, which is 
independent of time.  Correspondingly, the energy loss increases linearly
with length $L$ of the path, $\Delta E=\kappa_s\,L$.  The energy lost goes into
acceleration of the target quarks and production of new hadrons.

In the simplest version of the string model, where at most one string can
attach to a given quark, multiple interactions of the leading quark in the
nuclear medium are expected to have no influence on its energy loss
\cite{k90}.  Indeed, no matter what happens to the leading quark before the
hadronization is completed, it remains a color triplet and is slowed down
with the same density of energy loss $\kappa_s$ by the color-triplet string
attached to it:  the quark is always being retarded with the same $\kappa_s$.  
Although such a conclusion sounds puzzling, it does not mean that multiple
interactions in nuclear matter have no effect on hadronization.  These 
interactions do make the fragmentation function of the quark softer, 
{\it i.e.} more particles are produced in the fragmentation region of the 
nucleus ($dn/d\eta\propto A^{1/3}$) as the leading hadron diminishes its energy
\cite{kn1,kn2}.  This indeed may look like the result of energy loss, but
it happens because each rescattering of the color string in nuclear matter
initiates a new hadronization process from the very beginning, but with
decreased initial energy \cite{k90}.

The above treatment of the interaction is obviously oversimplified; in fact,
multiple interactions in nuclear medium do induce extra energy loss.  Indeed,
the the valence quarks are surrounded by parton clouds, only a part of
which can be resolved by a soft interaction.  Multiple interactions in
nuclear matter are obviously able to involve more of the partons.  
Correspondingly, the energy loss should grow more steeply than linearly in $L$.
 
This idea is realized in the dual parton model \cite{capella} (or quark-gluon 
string model \cite{kaidalov}) by assuming that each multiple interaction in 
the nuclear medium activates a new sea $\bar qq$ pair, leading to the 
formation of an extra couple of color strings.  The probabilities of multiple 
string production are given by the Abramovsky-Gribov-Kancheli (AGK) cutting 
rules \cite{agk}.  The dual parton model skips over the space-time development 
of string production and decay, which takes long time proportional to the 
initial energy, jumping directly to the final spectrum of produced particles.  
We, however, are interested in the early stage of hadronization when the 
projectile quark is still propagating through the nucleus and its cloud of sea 
partons is still coherent.  Therefore, we should follow the overall 
energy-loss of a nonperturbative constituent quark retarded by more than just 
one color string.

A quark covering the distance $z$ in the nucleus, from the point of the 
first inelastic interaction of the incident hadron, loses an energy per unit
length given by
\beq
\frac{d\,E}{d\,z}=
\kappa_s\Bigl[1 + \la n(z)\ra\Bigr]\ ,
\label{3.2}
\eeq
where $\la n(z)\ra = \sigma^{qN}\,\rho_A\,z$ is the mean number of collisions 
experienced by the quark over the distance $z$.  One can approximate the 
value of the nuclear density by $\rho_A\approx 0.16\,fm^{-3}$, since the path 
between the first inelastic collision and the DY pair production covers
essentially only the interior of the nucleus.  Correspondingly, the energy 
loss over a distance $L$ acquires a correction $\propto L$,
\beq
\Delta E = \kappa_s\left(L + {1\over2}\,
\sigma^{qN}\,\rho_A\,L^2\right)\ ,
\label{3.3}
\eeq

The phenomenological quark-nucleon cross section is model dependent.  In the 
additive quark model, $\sigma^{qN}=\sigma^{NN}_{in}/3\approx 10\,mb$.
The same value follows from the dual parton model \cite{capella,kaidalov},
which uses the weight factors $\sigma_n$ for the $n$-fold scattering appearing
in the Glauber model as powers of  $\sigma^{NN}_{in}$.  The new strings formed 
after each rescattering should be shared by three valence quarks.

In a more realistic model~\cite{k3p}, the main part of the hadronic cross
section corresponds to a soft nonperturbative interaction, which is unable to
resolve and free the gluons that are located with small transverse
separation $\la r_T\ra=r_0=0.3\,fm$ around the valence quarks \cite{kst2}.
This part of the cross section is independent of energy and does not obey
quark additivity.  Clearly, such soft interactions of the quark
skeletons of the proton contribute to the linear term in (\ref{3.3}).  
Only a semi-hard interaction is able to resolve the small-size gluon cloud
and a correspondingly rather small quark-nucleon cross section related to
the excitation of quark and gluon radiation. This process corresponds to
the higher terms in topological expansion in the dual parton model, {\it i.e.}
contributes to the $L^2$ term in ({\ref{3.3}). The predicted cross section
is $\sigma^{qN}=(9/4)\,r_0^2\,(s/s_0)^{\Delta}$, where \cite{k3p}
$\Delta=0.17$ and $s_0=30\,GeV^2$.  At $s=1600\,GeV^2$, one gets 
$\sigma^{qN}=4.2\,mb$, more than twice as small as the prediction of the 
additive quark model.

The mean values $\la L\ra$ and $\la L^2\ra$ in a heavy nucleus (with constant 
density) corrected for the mean free path $\lambda\approx 2\,fm$ of the 
incident proton in the nucleus are,
\beqn
\la L\ra&=&{2\over3}\,R_A-\lambda\nonumber\\
\la L^2\ra&=&{1\over2}\,R_A^2 -
{4\over3}\,R_A\,\lambda +
\lambda^2\ .
\label{3.4}
\eeqn
For example, for $R_A=6\,fm$ the nonlinear $L^2$ correction in (\ref{3.3}) is 
only $24\%$ and $10\%$ in the two models discussed above.  It decreases 
dramatically, of course, for lighter nuclei.  We neglect this small correction 
and assume for further applications that energy loss is a linear function of 
$L$.

\subsubsection{Perturbative QCD}\label{pqcd}

Although application of perturbative methods to soft processes is not
completely legitimate, one may hope to get at least the scale of the
effect by doing so.  Energy loss treated perturbatively originates from
gluon bremsstrahlung by a quark propagating through a medium. Here again
one should distinguish between two sources of energy loss, namely
gluon radiation originating in the first inelastic interaction, which
occurs even in the vacuum, and gluon radiation induced by multiple
interactions of the quark in the medium.

It was first demonstrated by Niedermayer~\cite{n} that as a consequence of 
coherence effects, (photon) gluon bremsstrahlung in the vacuum
carries energy away with a constant rate,
\beq
\Delta E(L)=E_q
\int d^2k_T\int\limits_0^{x_{max}}
dx\,x\,
\frac{d\,n_g}{dx\,dk_T^2}\ ,
\label {3.5}
\eeq
where $\vec k_T$ and $x$ are the transverse momentum and fraction of the 
quark light-cone momentum of the parent quark momentum carried  by the gluon,
respectively.  The upper integration limit $x_{max}$ is fixed by the condition
that all gluons contributing to $\Delta E$ are radiated within the path $L$. 
In order to be radiated, the gluon must lose coherence with the parent quark,
otherwise one cannot disentangle between the quark plus its color field and 
the quark plus the radiated gluon.  The distance $l_f$ over which this 
loss of coherence occurs and radiation formed 
is specified by the condition that the quark-gluon 
separation must exceed the transverse wave length of the gluon, 
which leads to 
$l_f=E_qx/k_T^2$, where $E_q$ is the quark energy \cite{miklos}. Note that 
this value is twice as small as the coherence length defined in (\ref{1}).
Applying the condition $l_f< L$ we get $x_{max}=Lk_T^2/E_q$.

The spectrum of radiated gluons with $k_T^2$ smaller than the mean momentum 
transfer squared in the collision was calculated in \cite{gb},
\beq
\frac{d\,n_g}{dx\,d^2k_T}=
\frac{3\,\alpha_s}{\pi^2\,x\,k_T^2}\ .
\label {3.6}
\eeq
Thus, the density of energy loss reads,
\beq
-\frac{d\,E}{d\,z}=\frac{3\,\alpha_s}
{\pi}\ \la k_T^2\ra\ ,
\label {3.7}
\eeq
and a quark loses energy in the vacuum with a constant rate proportional to 
$\la k_T^2\ra$. Note that this is not the induced energy loss in a medium.

The proper value of $\alpha_s\approx 0.4$ for gluon radiation
was calculated in \cite{k3p}. The transverse momentum squared of radiated 
gluons is rather large $k_T \approx 1/r_0=650\,MeV$ \cite{kst2,k3p}. 
With these values we evaluate the energy loss rate (\ref{3.7}) as 
$dE/dz\approx 0.8\,GeV/fm$, which is amazingly close to the string tension. 
Note that the radiative energy loss is not supposed to be an
alternative to what is given by the string model, but it is rather a 
different source that should be
added to it.  This is analogous to the additive contributions to the total
hadronic cross section  arising from nonperturbative strings and 
gluon radiation~\cite{k3p}

Quite a different situation arises if the quark originates from a hard
reaction as, say, in DIS or a high-$p_T$ scattering event, where the rate
of energy loss from gluon radiation is expected to be tremendous \cite{knp}.  

Another source of energy loss is induced gluon radiation due to multiple
quark interactions in the medium. This has received much attention recently
\cite{miklos,baier}, since it may serve as a probe for the production of dense
matter in relativistic heavy ion collisions \cite{baier}. The induced gluon
radiation was calculated perturbatively \cite{baier}, and it was concluded
that the energy loss grows like $\Delta E\propto L^2$. This is a direct
consequence of the relation (\ref{3.7}) between the rate of energy loss and
the transverse momentum squared of gluons, which follows that of the quarks
and is known to rise $\propto L$. The
relation between induced energy loss and nuclear broadening of the mean
transverse momentum squared of the quark was established in \cite{baier},
\beq \Delta E=\frac{3\,\alpha_s}{8}\, \Delta\la p_T^2\ra\,\la L\ra\ ,
\label {3.8} \eeq where the broadening of $\la p_T^2\ra$ was measured
\cite{peng} to be rather small $\Delta\la p_T^2\ra\approx 0.1\,GeV^2$ even
for tungsten. Therefore the coefficient of $\la L\ra$ in (\ref{3.8}) is
about $0.075\,GeV/fm$, an order of magnitude smaller than the string
tension $\kappa_s$.

The induced energy loss of a nonperturbative quark, both the string
and radiative parts, turn out to be a rather small fraction, less
than $\sim 10-15\,\%$, of the total energy loss even for heavy
nuclei.  Although this correction has a different $L$-dependence,
one can effectively absorb it into the main linear term.

The dominant constant parts of the energy loss rate $dE/dz$ (so-called vacuum 
energy loss), related to the string tension (Sect.~3.1.2) and to gluon 
radiation, have different origins and, as we have said, should be added.
The resulting rate of energy loss is thus expected to be
 \beq
-\frac{d\,E}{d\,z}\approx 2\,GeV/fm\ .
\label{3.9} 
\eeq

\subsection{The quark path in the nucleus}\label{path}

The DY reaction on a nuclear target is usually mediated by the debris
resulting from an inelastic collision between the incident hadron and a
bound nucleon on the front surface of a nucleus.  This debris, once
produced, propagates through the nucleus and produces the observed lepton
pair when it strikes a bound nucleon, as illustrated in Fig.~\ref{picture}
(top).
 \begin{figure}[tbh]
\includegraphics{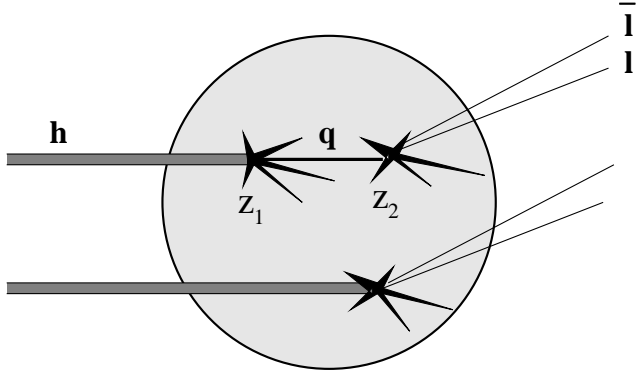}
\begin{center}
\vspace{5cm}
\parbox{13cm}
{\caption[Delta]
{\sl The space-time pattern for DY pair production off a
nucleus. The upper example illustrates a case when the
beam hadron experiences a soft inelastic interaction
prior the hard interaction in which the $\bar ll$ pair
is produced. The case where the DY pair is produced in the first
inelastic interaction is illustrated at the bottom.}
\label{picture}}
\end{center}
\end{figure}
One would not expect to observe any difference between this and the DY
reaction on a free nucleon for $z_2-z_1 \to 0$.  The reason is that the
primordial momentum distribution of the projectile partons cannot be
affected by soft inelastic interactions occuring over short times. However,
when the distance $z_2-z_1$ is finite, the soft projectile partons may lose
energy through hadronization and thus reach the point $z_2$ with
diminished energy.  However, at fixed $x_1$ the DY pair has to be produced 
with the same
measured momentum, {\it i.e.} with an increased fraction of the initial 
momentum.
As a result, the cross section for a DY pair produced with a given
longitudinal momentum turns out to be smaller on a nucleus than it is on a
free nucleon target.

It is usually incorrectly assumed that the quark propagates from the surface 
of the
nucleus to the point where the DY pair is produced, which would mean that
the mean quark path in the nucleus would be $\la L\ra\approx 3\,R_A/4$.  
According to Fig.~\ref{picture}, this should be shortened by at least the
mean free path of a proton in a nucleus, $2\,fm$.  This would substantially
reduce $\la L\ra$, by a factor of two or more, so that the mean path between 
the point of DY pair production and the
first inelastic interaction is actually shorter than the maximum possible
distance to the edge of the nucleus shown in Fig.~\ref{picture}.
Additionally, there is some probability (dominant for light and medium-heavy
nuclei) that the incident hadron has no interactions prior to DY pair
production at point $z_2$.  
In order to find the mean path length $\la L\ra$ of the projectile
quark in nuclear matter we should average
$(z_2-z_1)$ in accordance with Fig.~\ref{picture}, as done in~\cite{kn1},
\beqn
\la L\ra &=& (1-W_0)\,\frac{\sigma^{hN}_{in}}{A}
\int d^2b\,\int\limits_{-\infty}^{\infty} dz_2\,
\rho_A(b,z_2)\int\limits_{-\infty}^{z_2} dz_1\,
\rho_A(b,z_1)\,(z_2-z_1)\nonumber\\
&\times & \,
{\rm exp}\left[-\sigma^{hN}_{in}
\int\limits_{-\infty}^{z_1} dz\,\rho_A(b,z)\right]\ .
\label{2.0}
\eeqn
The exponential factor requires that there is no inelastic interaction
of the beam hadron prior point $z_1$. The probability of no 
inelastic interaction of the beam hadron in the nucleus prior
the DY reaction (see the bottom part of Fig.~\ref{picture})
$W_0$ reads,
\beq
W_0=\frac{1}{A\,\sigma_{in}^{hN}}\int d^2b\,\left[
1-e^{-\sigma_{in}^{hN}\,T(b)}\right] =
\frac{\sigma_{in}^{hA}}{A\,\sigma_{in}^{hN}}\ .
\label{2.0a}
\eeq

We calculated $\la L\ra$ for a
$pA$ collision taking $\sigma_{in}^{pN}= 30\,mb$ and using a Woods-Saxon
parameterization for the nuclear density \cite{jager}
 \beq
\rho_A(r)=\rho^0_A\,\left[1+
\exp\left(\frac{r-R_A}{c}\right)\right]^{-1}\ ,
\label{ws}
 \eeq
 where
 \beq
\rho_A^0=\frac{3A}{4\pi R_A^3}\left(1 + 
\frac{\pi^2c^2}{R_A^2}\right)^{-1}\ ,
\label{ws1}
 \eeq
and the nuclear radius $R_A$ and the edge thickness $c$ are fixed for each
nucleus by a fit to data on electron-nucleus scattering \cite{jager}.
  
As expected, $\la L\ra$ is much shorter than $3\,R_A/4$.  Both of these
results are plotted versus $A^{1/3}$ in Fig.~\ref{l}.
\begin{figure}[tbh]
\includegraphics{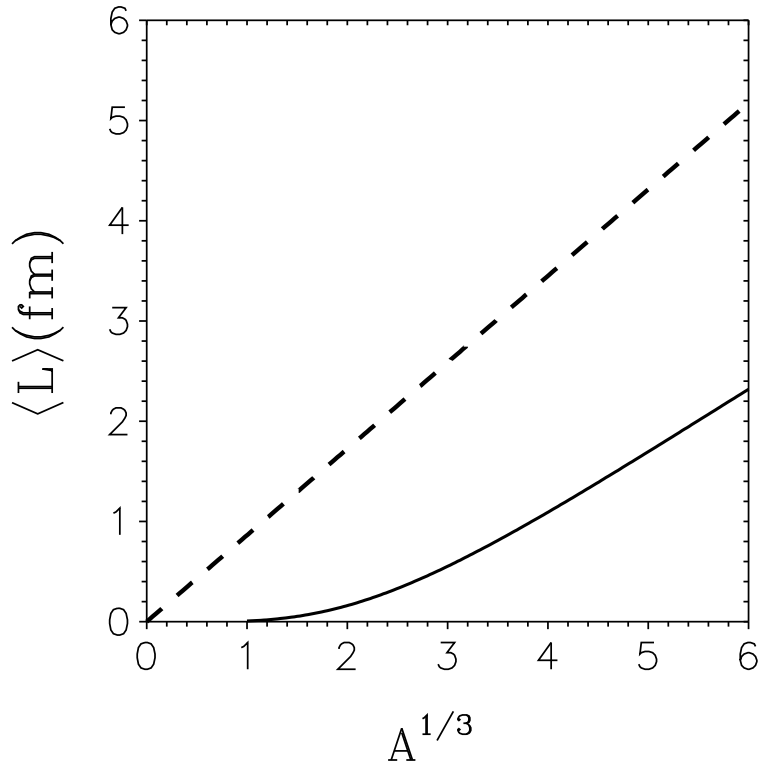}
\begin{center}
\vspace{10cm}
\parbox{13cm}
{\caption[Delta] {\sl
The mean length of the path of a quark between the
points of the first inelastic interaction of the beam
proton and DY pair production calculated with Eq.~(\ref{2.0})
as a function of $A^{1/3}$ (solid curve).
The dashed curve shows the usual expectation
$\la L\ra=3\,R_A/4$.}
\label{l}}
\end{center}
\end{figure}
The corresponding probability distribution in $L$ is given by the
expression
 \beq
W(L) = W_0\,\delta(L)\ +\ W_1(L)\ ,
\label{2.1}
 \eeq
where
\beqn
W_1(L) &=&
\frac{\sigma^{hN}_{in}}{A}
\int d^2b\,\int\limits_{-\infty}^{\infty} dz_2\,
\rho_A(b,z_2)\int\limits_{-\infty}^{z_2} dz_1\,
\rho_A(b,z_1)\,\delta(z_2-z_1-L)\nonumber\\
&\times& \,
{\rm exp}\left[-\sigma^{hN}_{in}
\int\limits_{-\infty}^{z_1} dz\,\rho_A(b,z)\right]\ ,
\label{2.2}
\eeqn
which is normalized as $\int_0^{\infty} dL\,W_1(L) = 1-W_0$.

Although they are not as large, the effects of energy loss affect the DY
cross section on a deuteron target as well. In this case the constant
$W^D_0$ is given by the standard Glauber formula \cite{glauber},
\beq
W^D_0 = 
1 - \frac{\sigma^{NN}_{tot}}{16\,\pi}
\int\limits_{0}^{\infty} dq^2\,
F_D(4q^2)\,e^{-B^{NN}_{el}q^2}\ ,
\label{2.2a}
\eeq
where $B^{NN}_{el}\approx 12\,GeV^{-2}$, and 
$F_D(q^2)$ is the deuteron charge formfactor parameterized and fitted in 
\cite{anisovich} to electron scattering data as
\beq
F_D(q^2) = c_1\,e^{-b_1\,q^2} +
c_2\,e^{-b_2\,q^2}\ .
\label{2.2b}
\eeq
Here $c_1=0.55$, $c_2=0.45$, $b_1=19.66\,GeV^{-2}$, and 
$b_2=4.67\,GeV^{-2}$.  This leads to $W_0^D=0.96$.
Correspondingly, the distribution function $W_1(L)$ 
for a deuteron is given by the wave function squared related to 
the formfactor (\ref{2.2b}),
\beqn
W^D_1(L) &=& (1-W_0^D)\,
\frac{L^2}{2 \sqrt{\pi}}\,\left[
\frac{c_1}{(4\,b_1+B^{NN}_{el})^{3/2}}\,
\exp\left(-\frac{L^2}{4\,(4\,b_1+B^{NN}_{el})}\right)\right.
\nonumber\\ &+& \left.
\frac{c_2}{(4\,b_2+B^{NN}_{el})^{3/2}}\,
\exp\left(-\frac{L^2}{4\,(4\,b_2+B^{NN}_{el})}\right)\right] .
\label{2.2c}
\eeqn

\subsection{Nuclear suppression caused by energy loss}
\label{ratio}

Since the DY cross section decreases steeply as $(1-x_1)^n$ for $x_1 \to
1$, any loss of energy by the quark by the time it reaches the point $z_2$
will result in a suppression of the DY cross section at large $x_1$. This
effect was first suggested and estimated in \cite{kn1}.  In the more
detailed approach taken here, we want to focus on the propagation and
energy loss of the projectile quark that takes part in the DY reaction.  In
this case, the ratio $R_{A/N}$ of the DY cross sections of a nucleus to 
a nucleon can be represented as 
 \beq R^{Eloss}_{A/N}(x_1) = W_0\ +\
\frac{\int\limits_{0}^{\infty} dL\,W_1(L) 
\int\limits_{(x_q)_{min}}^{1}
d\,x_q\,F^h_q(x_q)\,\, d\,
\sigma^{qN}_{DY}(\widetilde x_1^q)/ d\,
\widetilde x^q_1} {\int\limits_{x_1}^1 
d\,x_q\,F^h_q(x_q)\,\,
d\,\sigma^{qN}_{DY}(x^q_1)/d\,x^q_1} \ . 
\label{2.3} 
 \eeq 
Here $F^h_q(x_q)$ is the quark distribution function in the incident hadron;
$x_q$ and $x^q_1=x_1/x_q$ are the fraction of the light-cone momentum 
of the incoming hadron $h$ carried by the quark and the fraction of the quark 
momentum carried by the $\bar ll$ pair, respectively.  The lower integration 
limit is given by $(x_q)_{min}=x_1+\Delta E/E_h$, and $\widetilde
x^q_1=x_1/(x_q-\Delta E/E_h)$.  According to the results of Sect.~3.1 we
assume that the rate of energy loss $dE/dz=-\kappa$ is constant {\it i.e.}
$\Delta E=\kappa\,L$. Note that as the cross section 
$d\sigma^{qN}_{DY}(x^q_1)/dx^q_1$ in Eq.~(\ref{2.3}) corresponds to an
incident constituent quark, while that in (\ref{0.1}) corresponds to a
current (perturbative) quark.  To emphasize that these are different
quantities, we use variable $x^q_1$ in (\ref{2.3}) instead of $\alpha$  as in
(\ref{0.1}).
 
The energy loss $\Delta E$ varies due to fluctuations in the string
tension, the number and energy of radiated gluons, and due the to the
dependence of the distance $z_2-z_1$ shown in Fig.~\ref{picture} on the 
distribution $W(L)$.  We neglect the variation of $\kappa$, assuming 
$\kappa=const$, and integrate the numerator of (\ref{2.3}) over
$\Delta E$ weighted by the distribution $W(\Delta E)$, with $W(L)$ given in
(\ref{2.1}).

The first term on the {\it r.h.s.} of Eq.~(\ref{2.3}) corresponds
to the first term in (\ref{2.1}), which represents the probability of
no interaction preceding the DY reaction (the bottom part of
of Fig.~\ref{picture}).
The second term corresponds 
to one or more projectile interactions 
as illustrated in the upper part of Fig.~\ref{picture}.

\subsubsection{Distribution of the valence quarks in the proton}
\label{f_q}

As mentioned above, in the limit of short coherence length $l^{DY}_c \ll R_A$, 
the radiated photon arises from a hard DY fluctuation (one lasting for a 
short time) in the projectile quark, whereas the energy-loss mechanism is 
dominated by the softer fluctuations.  Therefore, to calculate (\ref{2.3}) we 
need to 
know the distribution function $F^h_q(x_q)$ and, correspondingly, the DY cross 
section $d\,\sigma^{qN}_{DY}(x^q_1)/d\,x^q_1$ for {\it soft} quarks.

One should rely on a phenomenology of soft hadronic interactions to find the 
soft quark distribution function $F^h_q(x_q)$. The proper approach is found
in the dual parton model \cite{capella} (quark-gluon string 
model~\cite{kaidalov}), which describes inelastic hadronic interactions
via convolution of the quark distribution functions with the fragmentation 
functions of quarks to hadrons. The success of this phenomenology in 
describing the available experimental data on soft hadronic interactions
justifies the quark distribution functions used above.

The proton quark distribution function in (\ref{2.3}) includes contributions 
from the valence {\it up} and {\it down} quarks,
\beq
F^p_q(x_q)= {8\over9}\,F^p_u(x_q)+
 {1\over9}\,F^p_d(x_q) \ ,
\label{2.4}
\eeq
where according to \cite{capella,kaidalov},
\beq
F^p_u(x_q)= N\ \frac{(1-x_q)^{1.5}}{\sqrt{x_q}}\,
\left[1+{6\over5}\ \frac{\sigma^{pN}_{tot}}{8\,\pi\,B^{pp}}\,
(1-x_q)\right]\ ,
\label{2.5}
\eeq
\beq
F^p_d(x_q)= N\ {6\over5}\ \frac{(1-x_q)^{2.5}}{\sqrt{x_q}}\,
\left[1+{8\over7}\,\frac{\sigma^{pN}_{tot}}{8\,\pi\,B^{pp}}\,
(1-x_q)\right]\ .
\label{2.6}
\eeq
Here $N$ is a normalization factor inessential for (\ref{2.3}); the factors 
$6/5$ and $8/7$ result from the normalization condition as well. The second 
terms in the square brackets in
(\ref{2.5}) and (\ref{2.6}) are the first unitarity corrections
(we neglect the small higher-order terms). They contain the
total $pp$ cross section $\sigma^{pp}_{tot}=40\,mb$, the
slope of elastic $pp$ scattering, $B^{pp}=12\,GeV^{-2}$,
and an extra factor of $2$ dictated by the AGK cutting rules \cite{agk}.
The behavior of these distribution functions as $x_q\to 0$ and
as $x_q\to 1$ is dictated by Regge phenomenology and is related to the
intercept of the leading meson trajectories $\alpha_R(0)=0.5$
and the nucleon trajectory $\alpha_N(0)=-0.5$ \cite{capella,kaidalov}.

\subsubsection{Quark-nucleon DY cross section}
\label{qN}

In order to evaluate $R^{eloss}_{A/N}(x_1)$ in Eq.~(\ref{2.3}) we first of
all have to fix the $x^q_1$ dependence of the quark-nucleon DY 
cross section.  As was emphasized above, in the limit of short coherence length
the quark that experiences initial state interactions is a constituent 
quark requiring the distribution function for soft quarks.  Therefore, the DY 
quark-nucleon cross section $d\sigma^{qN}_{DY}/dx^q_1$ should be treated 
correspondingly.  Its shape cannot be predicted reliably since it 
depends in an essential way on the model for constituent quarks,
which lies in the domain of nonperturbative effects.  Instead, we 
parameterize the 
cross section in the form,
\beq
\frac{d^2\,\sigma^{qN}_{DY}}{d\,M^2d\,x^q_1}
=K(M^2)\Bigl(1-x^q_1\Bigr)^m\ ,
\label{2.7}
\eeq
where $K(M^2)$ and $m$ are fitted parameters.  We use data \cite{pat} for the 
DY cross
section in $p+^2H$ collisions and fit it by the expression
\beq
\frac{d\,\sigma^{NN}_{DY}}
{d\,x_1} \propto
\int\limits_{x_1}^1
d\,x_q\,F^h_q(x_q)\,\,
\left(1-\frac{x_1}{x_q}\right)^m\ ,
\label{5.0}
\eeq
with $m=0.362 \pm 0.027$ taken to be independent of $M^2$.
An example of the fit to the E772 data on deuterium target is depicted in 
Fig.~\ref{deuteron} for the dimuon mass interval $7-8\,GeV$.
 \begin{figure}[tbh]
\includegraphics{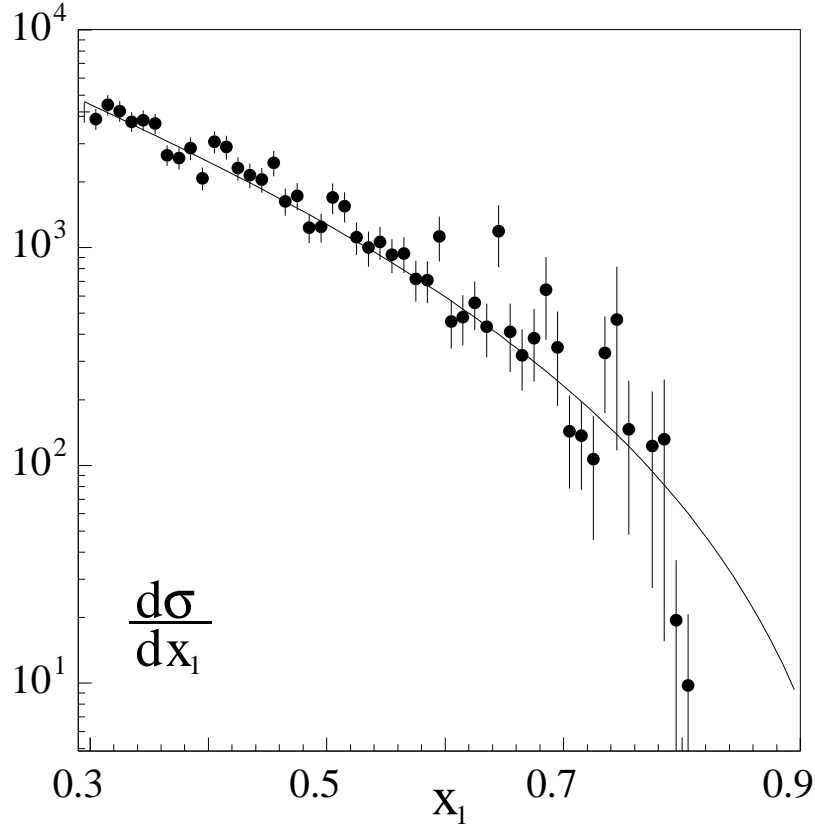}
\begin{center}
\vspace{11cm}
\parbox{13cm}
{\caption[shad1]
{\sl The cross section of the DY reaction on deuterium in arbitrary units
as function of $x_1$. The data are from the E772 experiment
for dimuon mass interval $7-8\,GeV$. The curve is the result of the fit
$\propto (1-x_1)^{0.362}$.}
\label{deuteron}}
\end{center}
\end{figure}

effects, 
%

\section{Nuclear shadowing: DY versus DIS}\label{shadowing}

In the target rest frame, it is clear that the longitudinal momentum transfer
to the nucleus in the DY reaction becomes small at high energies, in spite of
the large effective mass $M$ of the dilepton.  Therefore, different nucleons
start interfering (destructively) in the production of the DY lepton pair, a
phenomenon usually called shadowing.  

Naively, one would not expect significant shadowing for hard reactions such as 
DIS, DY processes, heavy flavor production, etc., because these processes have 
tiny cross sections and because bound nucleons clearly do not shadow each 
other.  However, substantial shadowing does exist for all of these,
particularly the DY process, at high energies.   The explanation is simple.  
First of all, shadowing arises from the soft components present in the hard 
reaction~\cite{hir}.  These soft components are small because the probability 
to develop a soft fluctuation in a hard reaction is small.  However, this small
size is compensated by a large interaction cross section.  Once a 
soft fluctuation
is created, the reaction is then driven by the subsequent hadronic 
interactions with the nucleus; these interactions are distinct from those 
producing the fluctuation to begin with and may be significant when the 
lifetime of the fluctuation (coherence time) is long.

The light-cone dipole representation of this reaction is especially
suitable for the calculation of nuclear shadowing.  The DY cross section
in this representation has the form \cite{hir,bhq,kst1},
 \beq
M^2\,\frac{d\sigma^{DY}_{NN}}
{dM^2\,dx_1} \propto \int\limits_{x_1}^1
\frac{d\alpha}{\alpha^2}\,
F^p_q\left(\frac{x_1}{\alpha}\right)\,
\int d^2r_T\,\left|\Psi_{\bar llq}(\alpha,\vec r_T)\right|^2\,
\sigma_{\bar qq}(\alpha r_T)\ ,
\label{4.1}
\eeq
where $\Psi_{\bar llq}(\alpha,\vec r_T)$ is defined in (\ref{7}).

If $l_c \gg R_A$, the $q\bar ll$ projectile fluctuations are frozen by
Lorentz time dilation during their propagation through the nucleus.
Therefore, these fluctuations are eigenstates of the interaction and
should have no inelastic corrections. One can simply eikonalize the dipole
cross section in (\ref{4.1}) in the case of a nuclear target
\cite{zkl,hir,kst1},
 \beq
\sigma_{\bar qq}(\alpha r_T)\,
\Rightarrow\, 2\int d^2b \left\{1\,-\,
{\rm exp}\left[-{1\over2}\,
\sigma_{\bar qq}(\alpha r_T)T(b)\right]
\right\}\ ,
\label{4.2}
\eeq
where the averaging over $\alpha$ and $\vec r_T$ is performed in accordance 
with (\ref{4.1}).  In the case of weak shadowing, one can expand the exponent,
dropping the higher-order shadowing corrections to obtain
\beq
R^{shad}_{A/N} = 
\frac{\sigma^{DY}_A}{A\,\sigma^{DY}_N} =
1\,-\,{1\over4}\,\sigma_{eff}\,\la T\ra \,+\,
O\bigl(\la\sigma^3\ra\bigr)\ ,
\label{4.3}
\eeq
where
 \beq
\sigma_{eff}(x_1,x_2,s)=\frac{\left\la
\sigma^2_{\bar qq}(\alpha r_T)\right\ra}
{\left\la
\sigma_{\bar qq}(\alpha r_T)\right\ra}\ ,
\label{4.3a}
\eeq
and where
\beq
\la T\ra =
{1\over A}\int d^2b\,T^2(b)
\label{4.4}
\eeq
is the mean value of the nuclear thickness function
 \beq
T(b) =
\int\limits_{-\infty}^{\infty} dz\,
\rho_A(b,z)\ ,
\label{4.5}
\eeq
with $\rho_A(r)$ the nuclear density distribution.

Note that naively one may expect a very small transverse size $\sim 1/M^2$
for the $|q\bar ll\ra$ fluctuations that define the dipole cross section
in (\ref{4.2}) and $\sigma_{eff}$.  Although these fluctuations appear with a 
vanishing probability at large $M$, soft fluctuations with $\alpha \to 1$ 
can have a large transverse separation \cite{hir,kst1}, 
[see (\ref{7}) - (\ref{8})]
 \beq
\la r_T^2\ra \sim \frac{1}{(1-\alpha)M^2}\ ,
\label{4.5a}
\eeq
as follows from (\ref{7}).  For this reason, these soft fluctuations have a 
large interaction cross cross section and can make a sizable contribution to 
$\sigma_{eff}$, in contrast to the harder ones~\cite{hir,kp}.

The transition region between the no-shadowing ($l_c \ll 2 fm$, or 
$x_2 \sim 0.1$) and asymptotic regimes ($l_c\gg \la L \ra$, or 
$1/x_2 \gg 3\,m_N\,R_A$), where Eq~(\ref{4.2}) is applicable,  is most 
complicated
and needs sophisticated calculations based on the path-integral approach
\cite{krt1,krt2}. Nevertheless, if shadowing corrections are small (which
is the case for the E772 and E866 data, where we face only the onset of
shadowing), Eq.~(\ref{4.3}) can be easily interpolated \cite{kp},
 \beq
R^{shad}_{A/N} \approx
1\,-\, {1\over4}\,\sigma_{eff}\,\la T\ra\,F_A^2(q_c)\ ,
\label{4.6}
\eeq
 where $F_A^2(q_c)$ is the longitudinal
formfactor of the nucleus, 
 \beq
F_A^2(q_c) = \frac{1}{\la T \ra}\,\int d^2b\,
\left|\int\limits_{-\infty}^{\infty} dz\,
e^{i\,q_c\,z}\,\rho_A(b,z)\right|^2\ .
\label{4.7}
 \eeq 
Strictly speaking, this formfactor must also be involved in the averaging 
procedure for
$\sigma_{eff}$ since $q_c=1/l_c$ depends on $\alpha$ and $k_T$.  In order
to speed up the fitting procedure (see below) we replace $l_c$ by its mean
value $\la l_c\ra$. Comparison with exact calculations for DIS done in
Ref.~\cite{krt2} demonstrates that this approximation is sufficiently
accurate.

Beryllium is too light for the Woods-Saxon parametrization (\ref{ws}) of
the nuclear density with which we describe heavier nuclei. Instead, we use for
beryllium the harmonic oscillator density \cite{jager},
 \beq
\rho_{Be}(r)=\rho_{Be}^0\,
\left(1+\alpha\,{r^2\over a^2}\right)\,
\exp\left(-\frac{r^2}{a^2}\right)\ ,
\label{4.7aa}
 \eeq
 where
 \beq
\rho_{Be}^0=\frac{9}{(\pi\,a^2)^{3/2}
\left(1+{3\over2}\alpha\right)}\ ,
\label{4.7ab}  
 \eeq
$a=1.77\,fm$, and $\alpha=0.631$.

The expression Eq.~(\ref{4.7}) is designed for medium and heavy nuclei and
cannot be applied to the deuteron target involved in our analysis. In this
case, one should use a different expression~\cite{glauber} [compare with
(\ref{2.2a})],
 \beq
R^{shad}_{D/N} = 
1 - \frac{\sigma_{eff}}{16\,\pi}
\int\limits_{0}^{\infty} dq_T^2\,
F_D(4q^2)\ ,
\label{4.7a}
\eeq
where $q^2=q_T^2 + q_c^2$, with $q_T$ the transverse momentum.  Using the
deuteron charge formfactor $F_D(q^2)$ in the of form Eq.~(\ref{2.2b}), we
arrive at shadowing for a deuteron,
 \beq
R^{shad}_{D/N} =
1 - \frac{\sigma_{eff}}{16\,\pi}\,
\left(\frac{c_1}{4\,b_1}\,e^{-4\,b_1\,q_c^2} +
\frac{c_2}{4\,b_2}\,e^{-4\,b_2\,q_c^2}\right)\ .
\label{4.7b}
\eeq

The effective cross section $\la\sigma^2\ra/\la\sigma\ra$ can be calculated
by averaging in accordance with (\ref{4.1}),
 \beq
\sigma_{eff}(x_1,x_2,s) =
\frac{\int_{x_1}^1
d\alpha\,F^p_q(x_1/\alpha)\,
[1+(1-\alpha)^2]\,\tau^2/\alpha^2\,
\int d^2r_T\,K_1^2(\tau r_T)\,
\sigma^2_{\bar qq}(\alpha r_T,x_2)}
{\int_{x_1}^1
d\alpha\,F^p_q(x_1/\alpha)\,
[1+(1-\alpha)^2]\,\tau^2/\alpha^2\,
\int d^2r_T\,K_1^2(\tau r_T)\,
\sigma_{\bar qq}(\alpha r_T,x_2)}\ ,
\label{4.9}
\eeq
 where $K_1(y)$ is the modified Bessel function and $\tau$ is defined in
Eq.~(\ref{8}). It is reasonable to use here the phenomenological
$x$-dependent dipole cross section \cite{gw} which describes very well data
from HERA for the proton structure function at high $Q^2$,
 \beq
\sigma_{\bar qq}(\rho,x) =
\sigma_0\left[1\,-\,{\rm exp}
\left(-\frac{\rho^2}{\rho_0^2(x)}
\right)\right]\ ,
\label{4.10}
\eeq
 where $\sigma_0=23.03\,mb$ and $\rho_0(x)=0.4\,(x/x_0)^{0.144}\,fm$,
$x_0=3.04\cdot 10^{-4}$. In the case of the DY reaction, $x=x_2$.

The results of the calculations for $\sigma_{eff}(x_1,x_2,s)$ at
$s=1600\,GeV^2$ are depicted in Fig.~\ref{eff-x2} as a function of $x_1$ for
different fixed values of $x_2$.
 \begin{figure}[tbh]
\includegraphics{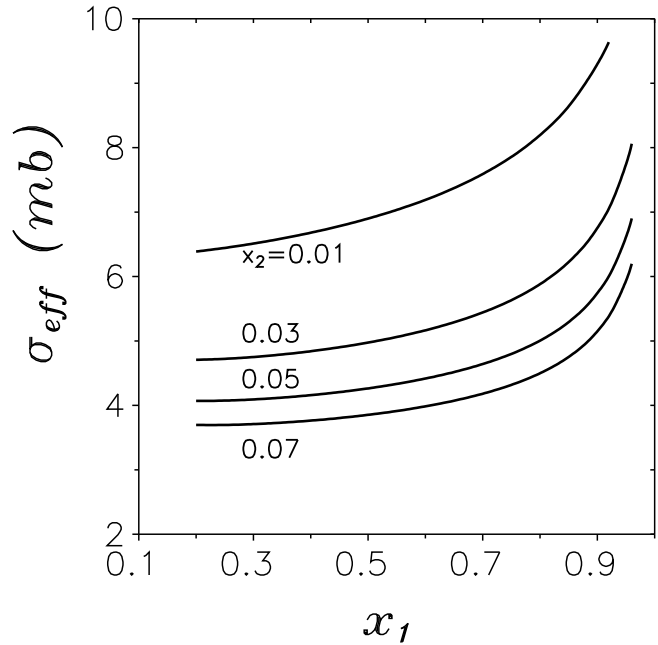}
\begin{center}
\vspace{8cm}
\parbox{13cm}
{\caption[Delta]
{\sl The effective cross section Eq.~\ref{4.9}) as a function of $x_1$ for 
different fixed
values of $x_2=0.02,\ 0.03,\ 0.05$ and $0.07$.}
\label{eff-x2}}
\end{center}
\end{figure}
Note that the effective absorption cross section $\sigma_{eff}$ substantially 
increases with $x_1$.  This is a manifestation of factorization breaking, 
since suppression ({\it i.e.} shadowing) of the nuclear structure function 
cannot increase with rising $M^2$ at fixed $x_2$. 
In the light-cone approach this behavior is easily interpreted:  the mean 
transverse separation squared that enters into (\ref{4.9}),
$\alpha^2\,\la r_T^2\ra \sim 1/\tau^2 \approx 1/[M^2\,(1-\alpha)]$,
increases as $x_1\to 1$ because $\alpha > x_1$.  Nevertheless, 
Fig.~\ref{eff-x2} shows that the $x_1$ dependence of $\sigma_{eff}$ 
is quite weak for $x_1<0.8$. 

Although we cannot test our calculations of shadowing by
comparing them to nuclear suppression in DY reactions, which are affected by 
energy loss effects, we may check them against nuclear shadowing in DIS.
In this case $\sigma_{eff}$ reads \cite{krt2},
\beq
\sigma^{DIS}_{eff}(x_2,Q^2) =
\frac{\int_0^1
d\alpha\,
[\alpha^2+(1-\alpha)^2]\,\epsilon^2\,
\int d^2r_T\,K_1^2(\epsilon r_T)\,
\sigma^2_{\bar qq}(r_T,x_2)}
{\int_{0}^1
d\alpha\,
[\alpha^2+(1-\alpha)^2]\,\epsilon^2\,
\int d^2r_T\,K_1^2(\epsilon r_T)\,
\sigma_{\bar qq}(r_T,x_2)}\ ,
\label{4.10a}
\eeq
where $\epsilon$ is introduced in (\ref{14}).

The calculated $\sigma^{DIS}_{eff}(x_2,Q^2)$ turns out to be rather
different from $\sigma^{DY}_{eff}(x_2,Q^2)$ calculated with
Eq.~(\ref{4.9}). Their ratio is plotted in Fig.~\ref{dis-dy}
as a function of $x_1$ for different dilepton masses. The fact that the
ratio is not unity is a deviation from factorization (except near 
$x_1 = 1$, where the factorization theorem does not apply since,
as we have noted, all
physics becomes soft here~\cite{bmht}).  However, the ratio
may be shown to approach unity logarithmically for large $M^2$
(and $x_1 < 1$), as required 
by the factorization theorem and as suggested 
by the results in the 
figure.
\begin{figure}[tbh]
\includegraphics{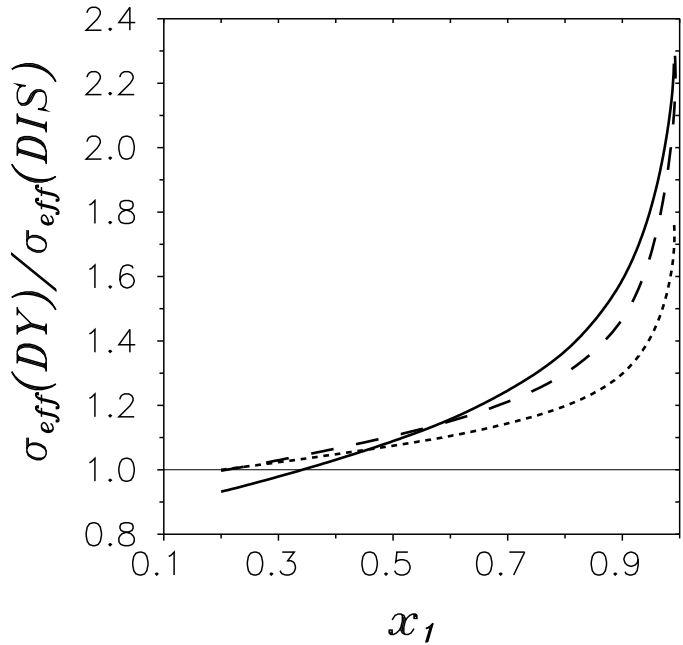}
\begin{center}
\vspace{8cm}
\parbox{13cm}
{\caption[shad1]
 {\sl Ratio $\sigma^{DY}_{eff}/\sigma^{DIS}_{eff}$ as a function of $x_1$ for
$M=4~(solid curve),\ 20$ (dashed) and $100\,GeV$ (dotted). Deviation from
unity demonstrates breakdown of QCD factorization (see text).}
 \label{dis-dy}}
\end{center}
\end{figure}

The effective absorption cross section controlling nuclear shadowing is
related to forward diffractive dissociation \cite{kk}. As a simple test,
we can compare our results for $\sigma^{DIS}_{eff}$ with data for
diffractive dissociation of highly virtual photons which were measured at
HERA to be about $10\%$ of the total DIS cross section,
 \beq
\frac{\sigma^{DIS}_{dd}}{\sigma^{DIS}_{tot}}=
\frac{\sigma_{eff}(x_1,x_2,s)}{16\,
\pi\,B^{DIS}_{dd}}\approx 0.1\ ,
\label{4.11}
 \eeq
 where the subscript $dd$ means diffractive dissociation, and
$B^{DIS}_{dd}\approx 5\,GeV^{-2}$. 

A more rigorous test would involve a direct comparison of the calculated nuclear
shadowing for DIS with data. Shadowing can be calculated using
Eqs.~(\ref{4.10a}) and (\ref{4.6}), where the argument of the formfactor,
$q_c=x\,m_N\la K^{DIS}\ra$, is given by Eq.~(\ref{12b}) (see
Fig.~\ref{lc-r2}). The results are plotted in Fig.~\ref{dis} as the
dashed curve and compared with data from the NMC experiment.
 \begin{figure}[tbh]
\includegraphics{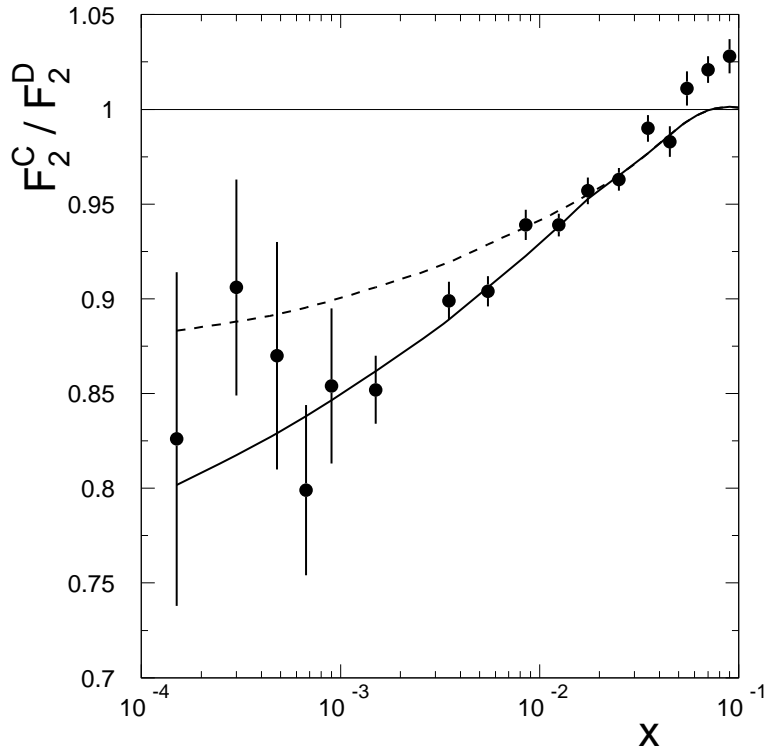}
\begin{center}
\vspace{10cm}
\parbox{13cm}
{\caption[shad1]
{\sl Normalized ratio of carbon to deuterium structure functions.
The data are from the NMC experiment \cite{nmc}.
The dashed curve is calculated using Eqs.~(\ref{4.6}),
(\ref{4.7a}), (\ref{4.10a}) for each data point
at the same values of $x$ and $Q^2$.
The solid curve includes also the effect of gluon shadowing as
it is calculated in \cite{kst2}.}
\label{dis}}
\end{center}
\end{figure}
Since the value of $Q^2$ in the data varies with $x$ we incorporate this
correlation in our calculations; we use the same set of $x$ and $Q^2$ as in 
the NMC data. Shadowing for the deuteron is also taken into account.  At large 
$x\sim 0.1$, the data display about $2\%$ antishadowing, the origin of which 
is poorly understood, although it is probably related to the same nuclear 
medium effects that cause the EMC effect.  As this dynamics is not included in 
our calculations, it is not surprising that we underestimate the data in this
region.

At smaller $x$ (and smaller $Q^2$), as expected, the calculated shadowing
keeps decreasing and does not seem to saturate.  This is due to the rise of
$\sigma_{eff}(x,Q^2)$ at small $x$ and $Q^2$.

At even smaller $x$, gluon shadowing also becomes important. The onset of gluon
shadowing occurs only at $x < 10^{-2}$ because it is related to
fluctuations containing extra gluons that are heavier than $\bar qq$
fluctuations.  Correspondingly, the coherence length for gluon shadowing is
shorter than it is for quarks \cite{krt2}. Nuclear shadowing for gluons at
small $Q^2$ relevant for the NMC data and at low $x$ is predicted in
\cite{kst2} (see Fig.~6 of that paper).  We multiply $F_2^C/F_2^D$ by the
gluon shadowing factor as calculated in \cite{kst2}. The result shown by
the solid curve is in good agreement with the data and confirms the
reliability of our description of shadowing, which contains no free
parameters.

\section{Energy loss versus shadowing}\label{eloss-shad}

By including shadowing explicitly in the analysis of the DY data, we 
are able to utilize the entire data set of the E772/E866 experiments for the 
ratio $R^{exp}(x,M^2)$.  This must be done recognizing that shadowing and 
energy loss
are complementary sources of nuclear suppression.  If the lifetime of a 
fluctuation containing the lepton pair substantially exceeds the size of the 
nucleus, we assume that the energy loss of the hadronic part of the
fluctuation does not affect the spectrum of dileptons, which are created as
a fluctuation long in advance of the nucleus\footnote{This is different
for nuclear broadening of the transverse momentum of a DY pair.  Since a 
nuclear target supplies a stronger kick to the quark part of the
fluctuation compared to a nucleon target, the nucleus is able to
break up the coherence of harder fluctuations, {\it i.e.} fluctuations with
larger transverse momentum of the lepton pair \cite{kst1}.}.  Thus, in the
two limiting cases of very long or very short coherence time one finds
either shadowing or energy loss present, correspondingly, not both
together.

The DY amplitude in the long $l_c$ limit differs from that in the short
$l_c$ limit by additional terms that account for interference between the
DY amplitudes involving different numbers of nucleons.  Within the
approximation that the double scattering correction dominates, which is
quite accurate if shadowing is small (appropriate for the kinematics of the
E772/E866 experiments), the interference term always appears multiplied by
the nuclear formfactor squared, as in Eq.~(\ref{4.6}). The transition from
the short $l_c$ regime, described by Eq.~(\ref{2.3}), to the regime of
$l_c\gg R_A$, described by Eq.~(\ref{4.6}), is likewise controlled by the
nuclear formfactor.  It is easy to write the interpolating expression,
namely
 \beq
R_{A/N}(x_1,M^2)=
\Bigl[1-F_A^2(q_c)\Bigr]\,
\Bigl[R_{A/N}^{Eloss}(x_1)-1\Bigr] +
R_{A/N}^{shad}(x_1,M^2)\ ,
\label{5.1}
\eeq
where $F_A^2(q_c)$ is given by (\ref{4.7}) and $R^{Eloss}_{A/N}$ and 
$R_{A/N}^{shad}$ are given by (\ref{2.3}) and 
(\ref{4.6}) respectively.  Obviously, (\ref{5.1}) correctly reproduces the 
short and long coherence length limits and mixes these effects
when both contribute. A nice feature of Eq.~(\ref{5.1})
is that the effect of energy loss is weakened 
with rising $l_c$ and eventually vanishes at $l_c\gg R_A$.

\subsection{Results}\label{fit}

With Eq.~(\ref{5.1}), we have adjusted $\kappa$ to fit the entire set of
ratios $C/D$, $Ca/D$, $Fe/D$, $W/D$, $Fe/Be$ and $W/Be$ from the 
E772~\cite{e772} and E866~\cite{vasiliev}  experiments double-binned in 
$x_1\geq 0.3$ and $M\geq 4\,GeV$. Since the normalization of the data is 
subject to 
systematic uncertainties, we included them in the fit by introducing
for each experiment 
an overall normalization $N=1 \pm \Delta N$ for the theoretical values of the 
nuclear ratios.  We allowed these to contribute to $\chi^2$ by treating
them as additional experimental points and assuming a normal distribution for 
them.  
A few examples of the fit results are depicted in 
Figs.~\ref{c-d}, \ref{w-d}, and \ref{w-be} (more bins in $M$ and a greater 
variety of nuclei were involved in the fitting).
\begin{figure}[tbh]
\includegraphics{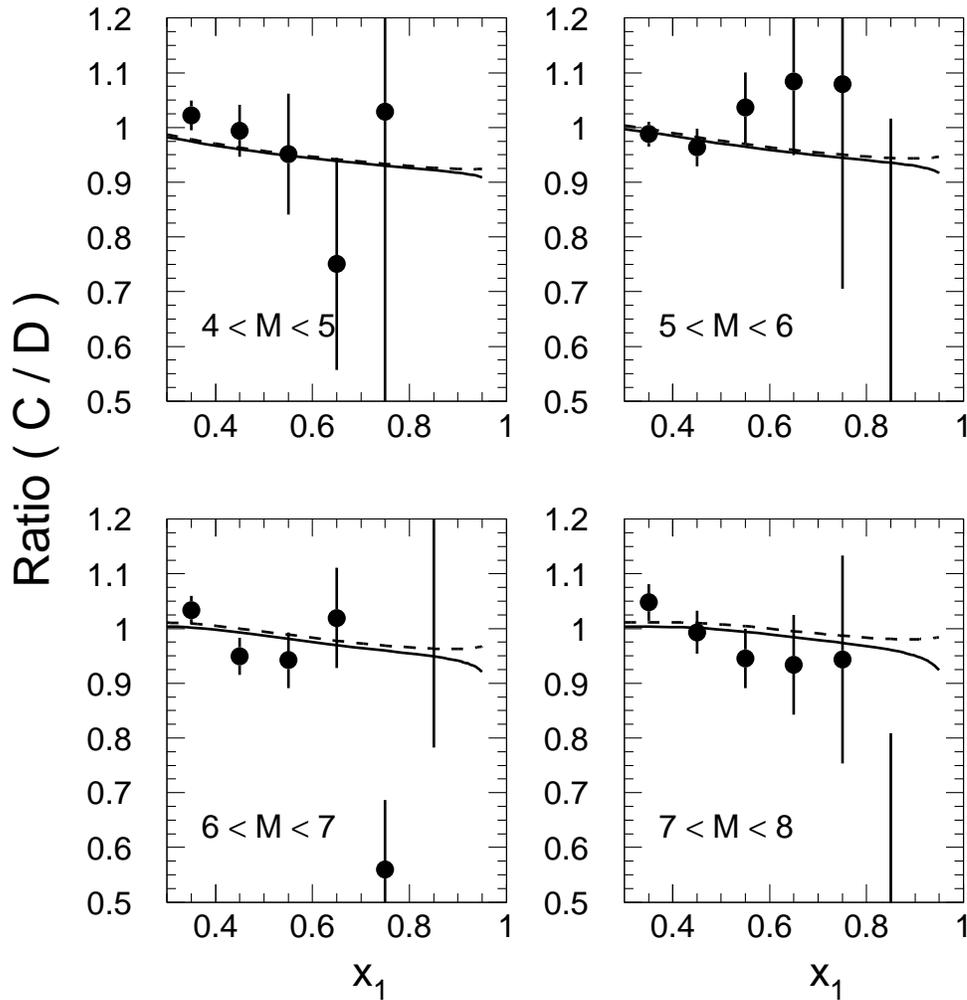}
\begin{center}
\vspace{13cm}
\parbox{13.3cm}
{\caption[shad1]
{\sl Examples of ratios of the DY cross sections on carbon to deuterium as 
functions of $x_1$ for various intervals of $M$.
Dashed curves correspond to net shadowing contribution,
solid curve show the full effect including shadowing and energy
loss.  Data from Refs.~\cite{e772},\cite{e772pc}.}
\label{c-d}}
\end{center}
 \end{figure}
 \begin{figure}[tbh]
\includegraphics{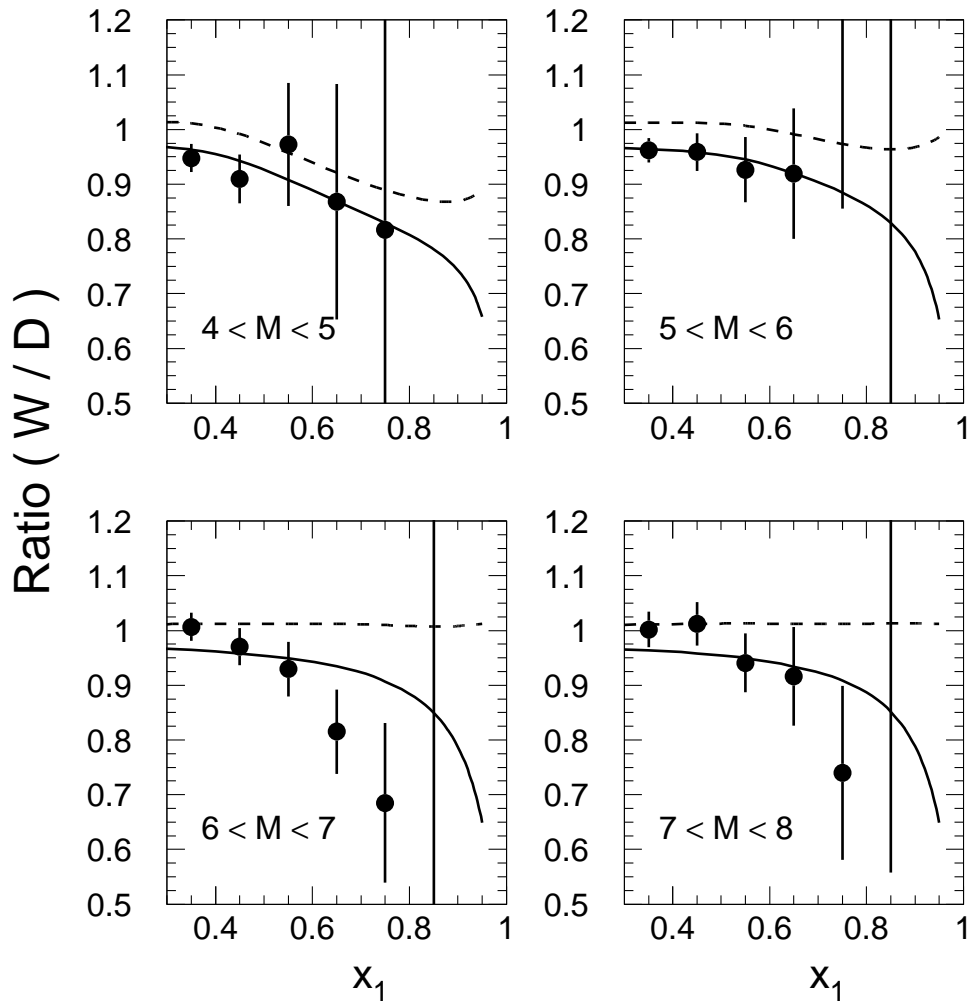}
\begin{center}
\vspace{13cm}
\parbox{13.3cm}
{\caption[shad1]
{\sl The same as Fig.~\ref{c-d}, except that the ratio
of tungsten to deuterium is shown.}
\label{w-d}}
\end{center}
 \end{figure}
 \begin{figure}[tbh]
\includegraphics{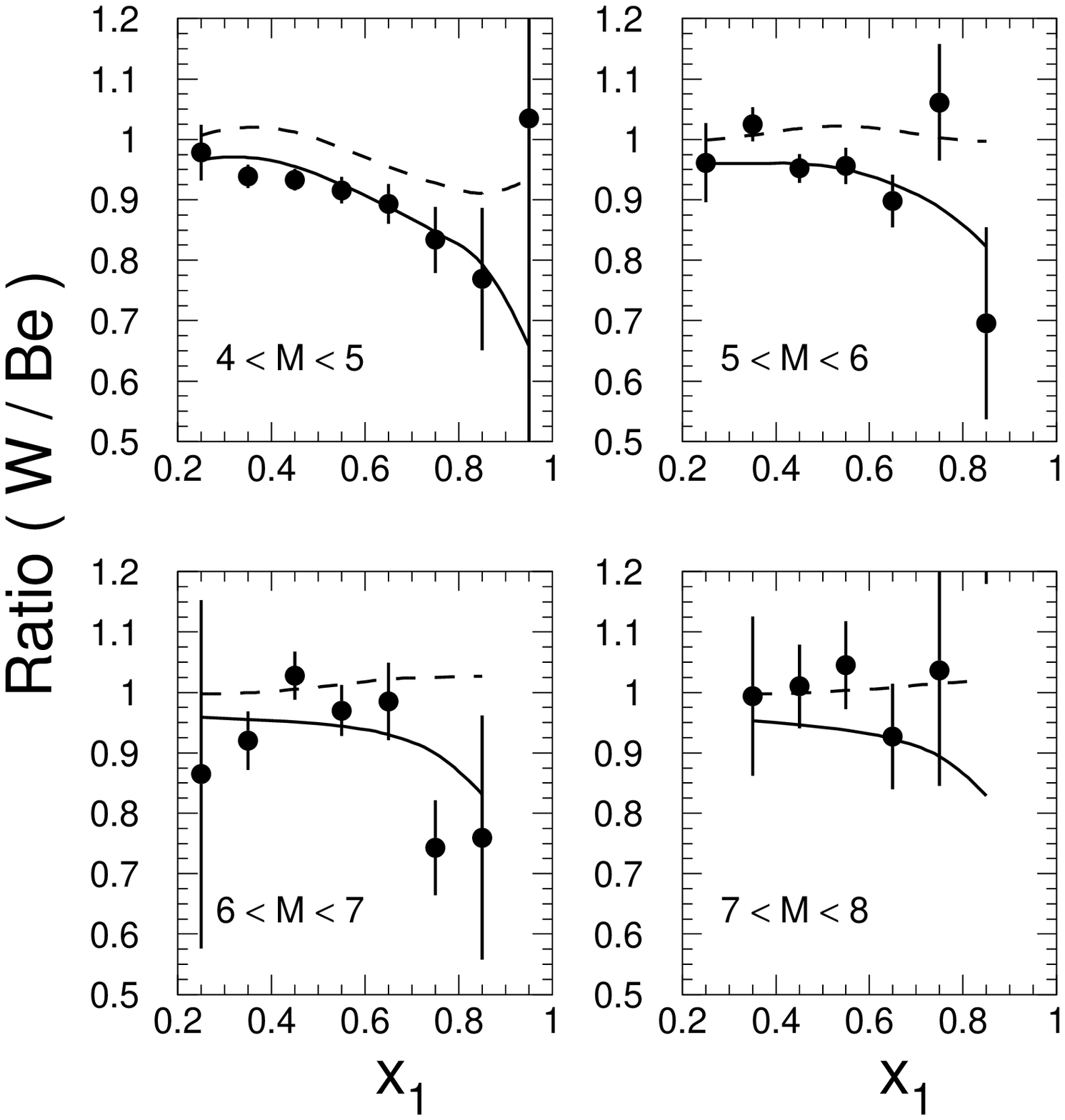}
\begin{center}
\vspace{13cm}
\parbox{13.3cm}
{\caption[shad1]
{\sl The same as Fig.~\ref{c-d}, except that the ratio
of tungsten to beryllium is shown, with data from Refs.~\cite{vasiliev},
\cite{e866pc}.}
\label{w-be}}
\end{center}
 \end{figure}
In addition to the solid curves, which show the full result of the
fit, we also show the contribution of shadowing alone in the dashed
curves.  Shadowing was calculated for the mean value of mass calculated
for each interval as $\sqrt{\la M^2\ra}$.

We found the following value for the rate of energy loss,
\beq
-\frac{d\,E}{d\,z} = 2.73 \pm 0.37\ GeV/fm\ ,
\label{5.2}
\eeq
with $\chi^2/d.o.f. = 0.9$, and normalization factors $N(E772)=1.014 \pm 0.005$
and $N(E866)=0.996 \pm 0.006$.

A nice feature of the fit is the consistency of the values of $dE/dz$
resulting from independent fits to the E722~\cite{e772pc} and
E866~\cite{e866pc} data,
 \beqn
-\frac{d\,E}{d\,z}\Bigr|_{E772} &=& 2.32 \pm 0.52\ GeV/fm\\
\label{5.3}
-\frac{d\,E}{d\,z}\Bigr|_{E866} &=& 3.14 \pm 0.53\ GeV/fm\ ,
\label{5.4}
\eeqn
with $\chi^2/d.o.f.=0.85$ and $\chi^2/d.o.f.=1.02$ for the E772 and E866 
data respectively, and normalization factors $N(E772)=1.01 \pm 0.006$
and $N(E866)=1.00 \pm 0.007$.

The rate of energy loss given in Eq.~(\ref{5.2}) is larger than the value
$1\,GeV/fm$ suggested by the string model.  However the string tension,
$\kappa_s=2\,\pi\,\alpha_R^{\prime}\approx 1\,GeV/fm$ is a static quantity,
related to the slope parameter $\alpha_R^{\prime}$ describing the orbital
excitation spectrum of hadrons.  Evidently the energy loss by a quark
includes an additional piece, such as the dynamical contribution arising
from gluon radiation \cite{kn1,kn2}.  The magnitude of the latter was
determined in section~\ref{pqcd} to be similar in size to the energy loss
arising from the formation of strings. Adding this contribution, the result
(\ref{5.2}) agrees with the expectation Eq.~(\ref{3.9}).

The normalization factors found for the E772 and E866 data are well within the 
quoted systematic
errors. These values of norms explain why the dashed curves representing
net shadowing in Figs.~\ref{c-d},~\ref{w-d} do not match unity for short
coherence length. At the same time the dashed curves slightly rise above 1
at large $x_1$ and $M$, demonstrating antishadowing. This is a result of
the delayed onset of shadowing toward small $x_2$ for heavy nuclei ($W$)
compared to light ones ($D$) as dictated by their formfactors.

One can see from Figs.~\ref{c-d}~-~\ref{w-be} that the effects of energy
loss and shadowing display quite different behaviors as a function of $A$,
$x_1$, and $M$.  For carbon nearly all nuclear suppression, to the extent
that it exists, comes from shadowing. This is because the mean path length 
available for energy loss in nuclear matter vanishes for nuclei as light as 
carbon, as one can see from Fig.~\ref{l}.  At the same time, for tungsten,
energy loss makes a substantial contribution to nuclear suppression, while 
shadowing vanishes for large masses because the nuclear formfactor 
falls steeply at large $x_2$.  This difference in the $A$ and $M$ dependence 
helps to disentangle the two effects rather effectively. 

Another specific signature of energy loss is a suppression which does not
vanish even at small $x_1$. Indeed, this effect of the order of $4\%$ seems
to be supported by data for tungsten in Figs.~\ref{w-d},~\ref{w-be}, while
data for carbon, Fig.~\ref{c-d} show no deviation from 1. This feature also
helps to single out the effect of energy loss since the data have
especially high accuracy in this region of small $x_1$.

Notice that the ratios of tungsten to beryllium shown in Fig.~\ref{w-be} reveal
an interesting feature; namely, at large dimuon masses this ratio rises with 
$x_1$ reaching values above one, reminiscent of antishadowing. Instead of this,
one might have expected a stronger shadowing for a heavier nucleus.  Of course 
Eq.~(\ref{4.6}) cannot lead to any antishadowing.  However, the onset of 
shadowing with decreasing $x_2$ must be delayed for heavier nuclei whose 
formfactors decrease more steeply with $q_c$.  We illustrate this feature in 
Figs.~\ref{shad-4.5} and \ref{shad-7.5}, corresponding to the dimuon masses 
$M=4.5$ and $7.5\,GeV$, respectively.
\begin{figure}[tbh]
\includegraphics{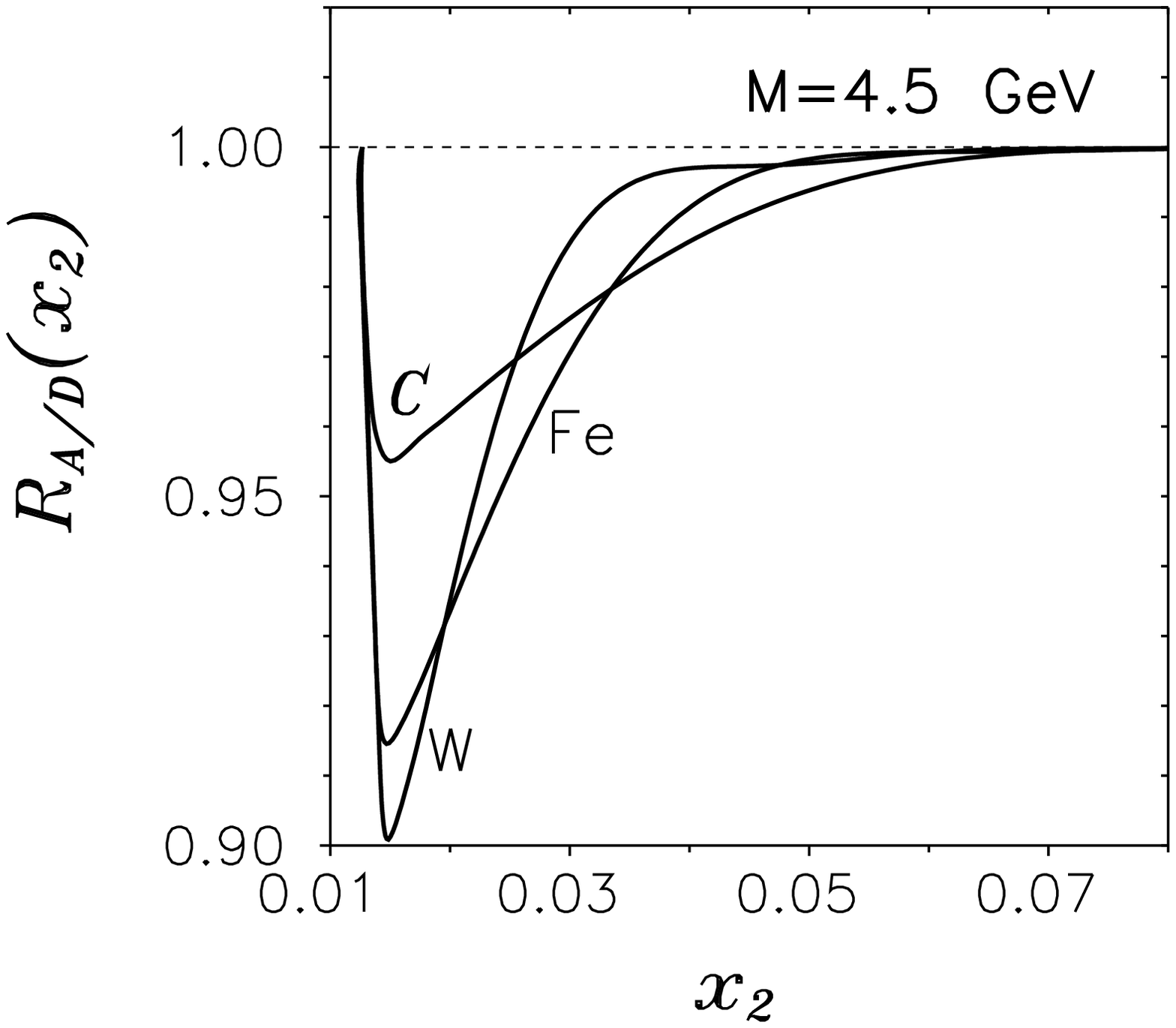}
\includegraphics{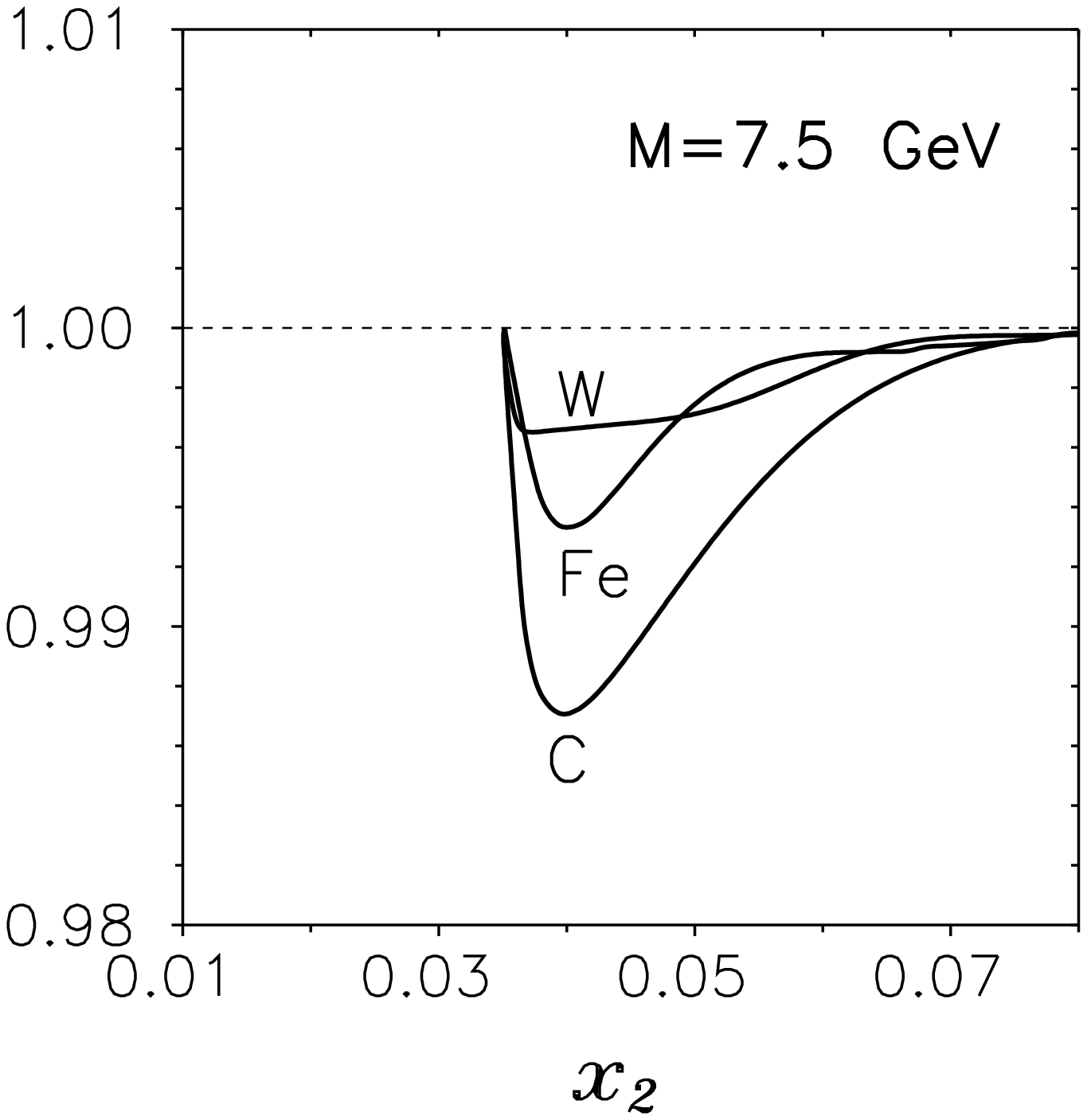}
\begin{center}
\vspace{7cm}
\parbox{13cm}
{\caption[Delta]
{\sl Shadowing in DY reaction on carbon, iron and tungsten as function
of $x_2$ at $M=4.5\,GeV$. Nuclear shadowing disappears
at large and small $x_2$ because the coherence length, Eqs.~(\ref{3}) and 
(\ref{4}), vanishes in these limits.} 
\label{shad-4.5}}
\parbox{13cm}
{\caption[Delta]
{\sl The same as in Fig.~\ref{shad-4.5}, but at $M=7.5\,GeV$.}
\label{shad-7.5}}
\end{center}
\end{figure}
We compare nuclear shadowing calculated with Eq.~(\ref{4.6}) for carbon, 
iron and tungsten as function of $x_2$. Indeed, one can see in 
Fig.~\ref{shad-4.5} that the onset of shadowing for heavier nuclei occurs
at smaller $x_2$ (compare with Figs.~5 and 6 in \cite{kst2}). 
At the larger dimuon mass of $M=7.5\,GeV$ in Fig.~\ref{shad-7.5}, the
behavior is seen to be even more complicated. In this case $l_c$ is so short 
that 
it suppresses the formfactor of tungsten at all $x_2$. In any case, these 
tiny variations of shadowing have no influence on our final results for 
energy loss.

Note that shadowing vanishes towards the kinematical limit $x_1=1$, i.e. 
minimal values of $x_2$. This is the result of the coherence length 
vanishing in this limit (see Fig.~\ref{lc-x2}). This property is irrelevant for
our analysis since there is no data in this region.

\subsection{Stability of the solution}\label{tests}

We have performed several tests of the stability of our results given in
(\ref{5.2}) - (\ref{5.4}).  First of all, as already mentioned, the analysis 
could be affected by any overlooked physical effects that are related to 
nuclear structure and that exist at large $x_2$ where no shadowing is 
expected.  To test this, we varied the relative number of shadowed and 
unshadowed events by imposing an upper cut-off $x_2^{max}$ on the values of 
$x_2$ allowed in the data set.  As this cut-off is lowered, more points are
affected by shadowing.  On the other hand, decreasing the cut-off $x_2^{max}$ 
diminishes the influence of any missed physical effects, as mentioned 
above.  We have
plotted the values of the energy-loss rate $\kappa$ resulting from separate
fits to the E772 and E866 data as a function of $x_2^{max}$ in Fig.~\ref{x2max}.
 \begin{figure}[tbh]
\includegraphics{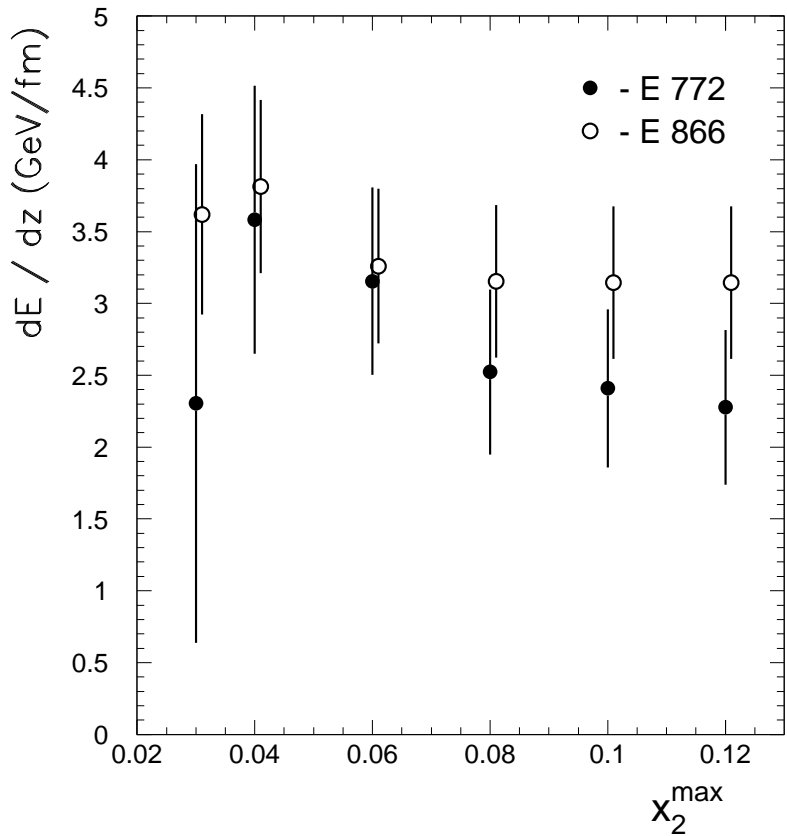}
\begin{center}
\vspace{11cm}
\parbox{13cm}
{\caption[x2max]
{\sl The rate of energy loss $-dE/dz=\kappa$
as a function of the upper cut-off $x_2 < x_2^{max}$. 
Closed and open circles correspond to fits to the E772
and E866 data respectively.}
\label{x2max}}
\end{center}
 \end{figure}
The results of the fits to both sets of data appear reasonably stable,
confirming the correctness of our calculations for nuclear shadowing of
the DY cross section.

Secondly, we tested the sensitivity of our results to the magnitude of the
shadowing in the analysis. We replaced $1-R^{shad}_{A/D} \Rightarrow
C_{shad}(1-R^{shad}_{A/D}$), where the factor $C_{shad}$ was varied.
Eliminating shadowing entirely by fixing $C_{shad}=0$, we found
$-dE/dz=2.34\pm0.37\,GeV/fm$ with $\chi^2=240$ from the common fit of the
E772 and E866 data. Next, making another extreme assumption,
we doubled the amount of shadowing by fixing $C_{shad}=2$ and found
$-dE/dz=3.00\pm0.37\,GeV/fm$ with $\chi^2=259$. Whereas one might be tempted
to conclude from this exercise that the treatment of shadowing is unimportant 
the extraction of energy loss, one clearly sees from from Fig.~\ref{x2max} 
that this is not true:  the energy-loss value determined by the data acquires 
important contributions from the shadowing region.  We also checked the
optimal value of $C_{shad}$ by treating it as a free parameter and found
$C_{shad}=0.88 \pm 0.18$ with $\chi^2=218$, which confirms correctness
of our calculations for shadowing. Thus, we conclude that by varying the amount
of shadowing over a wide range, the rate of energy loss remains unchanged
within error bars, while $\chi^2$ becomes substantially worse.


Next, we examined the accuracy of the double-scattering approximation
in Eq.~(\ref{4.6}). We had assumed that if the shadowing effect is weak, we
could neglect the triple and higher order rescattering terms. The exact
calculation of these corrections in \cite{krt2,krt3} is too
cumbersome to be used in the fitting procedure. Instead, one can
include the multiple interaction corrections in the same way as in
Ref.~\cite{kk}, {\it i.e.} assuming that rescatterings happen with the same 
cross section $\sigma_{eff}$ that governs the shadowing correction in
(\ref{4.6}). Then, the formfactor Eq.~(\ref{4.7}) should be modified as,
 \beq
\tilde F_A^2(q_c) = \frac{2}{\la T \ra}\int d^2b
\int\limits_{-\infty}^{\infty} dz_2\,\rho_A(b,z_2)
\int\limits_{-\infty}^{z_2} dz_1\,\rho_A(b,z_1)\,
\exp\left[iq_c(z_2-z_1) - 
\frac{\sigma_{eff}}{2}
\int\limits_{z_1}^{z_2} dz\,\rho_A(b,z)\right]\ .
\label{5.5}
 \eeq 
We repeated the fit replacing $F_A^2(q_c)\Rightarrow \tilde
F_A^2(q_c)$ in Eqs.~(\ref{4.6}) and (\ref{5.1}) and found no visible
deviation from the result in Eq.~(\ref{5.2}).

It was also of interest to know whether our results would change if,
instead of using Eq.~(\ref{5.1}) to combine the effects of shadowing and
energy loss, we make the maximal-strength mixture of these two effects as
assumed in \cite{vasiliev}; namely, $R_{A/N} = R^{Eloss}_{A/N}\,
R^{shad}_{A/N}$.  Fitting all data when we do this, we find nearly the same
value of the energy loss rate that we previously obtained,
$-dE/dz=2.18\pm0.31\,GeV/fm$.

One can also extract from the results of these modifications a scale for
the model dependence of the systematics in the rate of energy loss. We 
evaluate such a systematic error at about $\pm 0.5\,GeV/fm$.
Note that although all these variations of the fitting
procedure do not affect the results very much, they do substantially increase 
the value of $\chi^2$ compared to the fit done in Sect.~\ref{fit}.

The pleasing stability of the results confirms the conclusion of  
Sect.~\ref{fit} that there should be no confusion between the effects of 
shadowing and energy loss (provided that the data 
are sufficiently exact and binned both in $x_1$ and $M$) since these two 
effects behave very differently as functions of $A$, $x_1$ and $M$.

Finally, we would like to call attention to the fact that the data we are
analyzing for $\kappa$ cover the kinematical region of the target $x$
extending all the way out to large values of $x_2$. This is a concern,
since for $x_2>0.1$ there exist EMC and antishadowing effects in nuclear
quark distribution function that we have not yet accounted for in the
theory.  We suppress  the contribution of large $x_2$ selecting 
data with $x_1>0.3$. Besides, we argue next that these effects 
make a negligible contribution to the data we are analyzing.

To show this, we first remark that the EMC/antishadowing effects appear
in the {\it quark} distribution function of the nuclear target,
corresponding to antiquarks in the proton projectile.  We expect that the
EMC/antishadowing corrections, if appreciable, can affect the analysis
only at small $x_1$, since antiquarks in the beam contribute only for
small $x_1$.  These corrections enter the theory as expressed in
Eq.~(\ref{5.1}) in both $R^{shad}_{A/N}(x_1,M^2)$ (Eq.~(\ref{4.6})) and
$R^{Eloss}_{A/N}(x_1)$ (Eq.~(\ref{2.3})).

Consider first the shadowing contributions.  Shadowing is appreciable
only for very small $x_2<0.03$.  Taking $s=1600~GeV^2$, these small
values of $x_2$ correspond to values of $x_1=M^2(s x_2)^{-1}>0.3$ at the
smallest values of $M$ in our data set.  For $x_1>0.3$, the antiquark
distribution function is quite small, and in any case $\sigma_{eff}$,
Eq.~(\ref{4.9}), is not very sensitive to how we average.  So, we can
neglect the antiquarks in the shadowing region.

Consider next the energy-loss contribution.  For E772, all points in the
data set correspond to $x_1>0.3$.  This means that the data points are
uncontaminated by EMC and antishadowing effects ({\it i.e.}, $x_2<0.1$)  
for values of $M<6.9~GeV$.  However, for larger values of $M$ in E772,
and for all values of $M$ in E866, some region of $x_1$ will be affected.  
We have made quantitative estimates of the size of this contamination by
comparing the quark and antiquark contributions to the integral in
Eq.~(\ref{2.3}).  We evaluated the antiquark contribution using the same
the cross section $d\sigma^{qN}_{DY}(x^q_1)/dx^q_1$ that we found for an
incident quark, Eq.~(\ref{2.7}), and using the simple
expression~\cite{rpp} \beq {\bar q}(x_1)={0.7(1-x)^8\over 4x}
\label{qbar} \eeq for the atiquark distribution.  For $A=184$ and
$\kappa=2.28 $ GeV/fm, we find that the antiquark contribution is only
0.3\% of the quark contribution at $x_1=0.1$. Additionally, the relative
size drops rapidly for larger values of $x_1$, allowing us to conclude
that the EMC/antishadowing effects make a tiny contribution to the DY
cross section throughout the kinematical region we are considering and
justifying our neglect of these effects in the data analysis.

\subsection{Why a previous analysis~\cite{vasiliev} found no energy
loss}

One may wonder why the rate of energy loss $-dE/dz$ deduced from the E866 data 
in a previous analysis~\cite{vasiliev} is so much smaller 
($-dE/dz < 0.44\,GeV/fm$) than the value we found fitting the same data.  In 
this section we address this question.~\footnote{We compare only with the
second of three scenarios for energy loss considered in \cite{vasiliev}. The
first one (energy loss is proportional to the initial energy)  violates the
Landau-Pomeranchuk principle (see discussion in \cite{bh}), while the third
version (induced energy loss)  corresponds to energy loss induced by multiple
interactions of a quark propagating through infinite nuclear matter
\cite{baier}.  However the dominant contribution comes from vacuum energy loss
(see Sect.~\ref{models}) which has a constant and energy independent rate and is
properly treated with the second version considered in \cite{vasiliev}.}
A value smaller than the string tension $\sim 1\,GeV/fm$ would be at the same 
time both surprising and a serious problem for the theory.

One obvious reason for a difference is the effect discussed in Sect.~\ref{path} 
and Fig.~\ref{l}, namely the observation that when the multiple scattering
of the incident proton is properly considered, the mean path of a quark
available for vacuum energy loss in nuclear matter is more than twice as short 
as it would be if such scattering were neglected.  Taking this effect alone,
we estimate that it would increase the 1$\sigma$ upper limit for the rate of 
energy loss found in Ref.~\cite{vasiliev} to $-dE/dz < 1\,GeV/fm$.  

However, there are deeper reasons for the differences between the results of
our analysis and that in Ref.~\cite{vasiliev}, which treats the DY process as 
$q\bar q$ annihilation as in the leading-order parton model.
These reasons are related to the fact that the parton
model interpretation of the space-time development of the interaction is
not Lorentz invariant and depends on the reference frame.  As mentioned
already in Sect.~2.1, the lack of Lorentz 
invariance arises because objects 
such as partons are quantum fluctuations rather than classical particles.
Because of their quantum nature, the partons live and die differently in the 
different frames.  Thus, one must avoid trying to look at the same parton in 
different reference frames, even though $x_1$ is invariant under relative 
Lorentz boosts.  For example, if the DY process is treated in the lab frame, 
then the fraction of light-cone momentum carried by the valence quark, $x_q$, 
is larger than $x_1$ [see Eq.~(\ref{5})].  However, in the dilepton 
rest frame $x_q=x_1$; this is no surprise since they are not the same parton!

\subsubsection{Advantages of Target Rest Frame Formulation}\label{trf}

From the above discussion, it is clear that a  consistent approach to the
calculation of energy loss would, in principle, have to treat all effects
in the same frame.  In the case of the standard parton model
approach~\cite{dy}, this would be the rest frame of the dilepton.  In this
frame, the moving nucleus is
subject to a contraction by the Lorentz $\gamma$-factor, $R^*_A=R_A/\gamma$,
where
 \beq
\gamma = \frac{M}{2\,x_2\,m_N}\ .
\label{6.1}
 \eeq
The energy loss $\Delta E^*$ in this reference frame is diminished by the same
factor compared to the one in the nuclear rest frame,
$\Delta E^*=\Delta E/\gamma$.  Indeed, the vacuum energy loss is medium
independent and proportional to $L^*=L/\gamma$. The induced energy loss is
proportional to the Lorentz boosted nuclear density $\rho_A^*=\gamma\,\rho_A$,
but also to $(L^*)^2=L^2/\gamma^2$.  At the same time, the cloud of antiquarks
in the target nucleon spreads in longitudinal direction over distance
$\sim 2/M$. Thus at small $x_2$,
 \beq                      
x_2 \ll \frac{1}{m_NR_A}\ ,
\label{6.2}
 \eeq
(this is exactly the condition for long coherence length, {\it i.e.} for
shadowing) the size of the antiquark cloud substantially exceeds the
nuclear radius.
                                      
Unfortunately, in the rest frame of the dilepton the position of
$\bar qq$ annihilation cannot be localized to better than $\sim 1/M$,
while the nucleus is squeezed by the Lorentz transformation to a pancake
shape of a much smaller thickness $\sim x_2 R_A m_N/M$.  The annihilation
process clearly lacks sufficient resolution to probe such small
distances.  Additionally, one cannot say whether it is the quark or the
antiquark that propagates through the nucleus prior to annihilation, a
distinction that obviously must be made even before considering the issue    
of energy loss.                                                             
                                                                             
For these reasons, it is quite tempting to switch reference frames           
(which, however, we have argued is invalid, unless it is done properly) and
view the reaction from the rest frame of the nucleus.  Just such a switch
occurred in Refs.~\cite{gm,vasiliev}.  For example, in Ref.~\cite{vasiliev}
the shift $\Delta x_1$ caused by
energy loss was first fit to the E866 data in the rest frame of the
dilepton.  The authors then switched to the rest frame of the nucleus,
writing their Eq.~(2) in this frame and assuming, in accordance
with~\cite{n,bh}, that $\kappa_2$ is a constant independent of $x_1$,
$x_2$ and $s$.  They find the rate of energy loss to be $\Delta E=\Delta
x_1 E/(3R_A/4)$, where $E=800 GeV$ is the beam energy.
                
One might try to avoid these inconsistencies by treating the DY reaction as
$q\bar q$ annihilation in the rest frame of the nucleus, imagining the
projectile parton traveling through nuclear matter and suddenly annihilating
with a sea antiquark in one of the bound nucleons.  This is, however,
problematic for                                                             
several reasons.  First of all, the sea antiquarks available for such        
annihilation are defined only in the infinite momentum frame (this is why    
Weitz\"acker-Williams photons do not exist in the rest frame of the      
electron--they are its static field). Secondly, annihilation of a high-energy
massless quark with an antiquark at rest to a dilepton of mass $M$ violates
energy-momentum conservation. To fix this problem, one may introduce
next-to-leading order corrections, {\it i.e.} gluon radiation.  As some of
these gluons are radiated in advance of the dilepton and some later, one
cannot treat the DY reaction as instantaneous but rather must consider its
space-time development.  Unavoidably, one arrives at the picture employed in
present paper, where the DY pair is treated on the same footing as the gluons,
{\it i.e.} as bremsstrahlung.  

One is thus led in a natural way to the target
rest frame formulation, which clearly
identifies the incident quark as the one that propagates through the nucleus
and loses energy prior to the radiation of the ${\bar ll}$ pair.
The relative contribution of shadowing and this energy loss is governed by the
coherence length appearing as an argument of the longitudinal form factor in 
Eq.~(\ref{5.1}).  Specifying this dependence is essential for the 
determination of the correct 
value of the rate of energy loss from the data.  This is because many of the 
E866 data are located at small $x_2$, where the coherence length is relatively 
long (Fig.(\ref{lc-x2})), implying that shadowing dominates over energy loss in 
all but the heaviest nuclei.  Ignoring this dependence on $x_2$ would thus 
tend to overemphasize shadowing and lead to a diminished rate of energy loss, 
just as in Ref.~\cite{vasiliev}.

\subsubsection{Comparison of Results for Shadowing}

Still a substantial fraction of the E866 data is located at rather large $x_2$
where no shadowing is expected.  For these events, another difference exists
between Ref.~\cite{vasiliev} and our results.  This can be traced to the fact 
that the shadowing correction in \cite{vasiliev}, which is taken from the 
phenomenological analysis by Eskola, et al.~\cite{eks}, is quite 
different from ours in this region.  We compare our shadowing calculated for 
tungsten at $M=4.5$ depicted in Fig.~\ref{eks} by the dashed curve, with the 
one from Ref.~\cite{eks} (solid thin curve) that was used in 
Ref.~\cite{vasiliev} to correct the E866 data.  One can see quite clearly
the difference between these two prescriptions.
 \begin{figure}[tbh]
\includegraphics{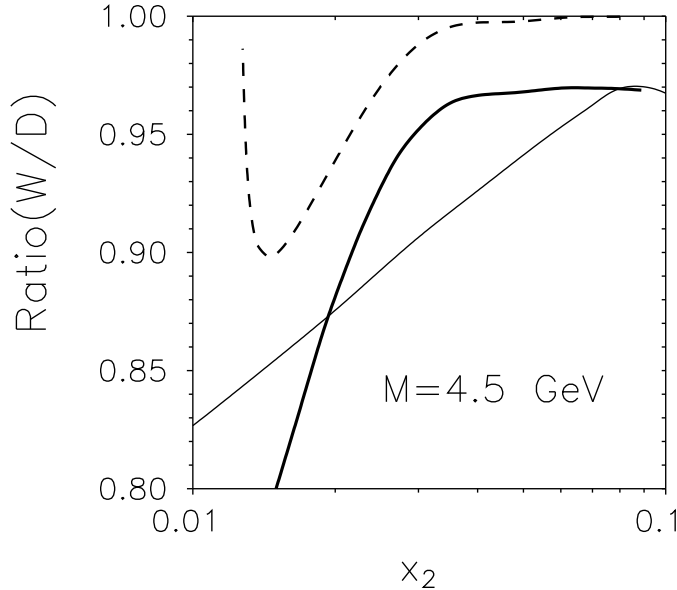}
\begin{center}
\vspace{7.5cm}
\parbox{13cm}
{\caption[x2max]
{\sl Comparison of nuclear shadowing (dashed curve) and combined effect of 
shadowing and energy loss (thick solid curve) presented in our analysis, 
with ``shadowing'' for antiquarks found in \cite{eks} from the combined 
analysis of DIS and DY data (thin solid curve).  All calculations are done 
for tungsten and $M=4.5\,GeV$. The behavior of the dashed curve at small
$x_2$ is explained in the text.}
\label{eks}}
\end{center}
 \end{figure}

The source of the difference at large $x_2$ arose, we believe, from 
a confusion between energy-loss effects and shadowing in the 
phenomenological extraction of the antiquark densities in Ref.~\cite{eks}.
Shadowing for antiquarks cannot be extracted in a model independent way
solely from DIS data, which are blind to the sign of the electric charge.
This is why the E772 data for the DY reaction were included in the fit of 
Ref.~\cite{eks} together with a variety of DIS data.  Assuming QCD 
factorization, the evolution equations were applied to the parton 
distributions in nuclei. As usual, one needs to know the input 
$x$-distributions at a medium high scale $Q^2_0$.  The shapes of these 
distribution were guessed in 
Ref.~\cite{eks}, and then their magnitude 
was fit to data.  No physical input beyond the QCD evolution was 
incorporated in the analysis.  It is not a 
surprise that substantial ``shadowing'' effects were found for antiquarks even 
at large $x_2$ where no quantum interference is possible because of the 
shortness of the coherence length. It was assumed in Ref.~\cite{eks} that the 
ratio of the antiquark densities is constant for $x > 0.08$.  The magnitude of 
this ratio was fit to the E772 data and found to be less than 1, as one can see 
from Fig.~\ref{eks}.  We think that this is where the misinterpretation of 
the energy loss effects as shadowing occurred in Ref.~\cite{eks}. Indeed, as we 
emphasized in Sect.~\ref{fit} and one can see in Figs.~\ref{c-d}~-~\ref{w-be},
energy loss suppresses the DY nuclear cross section even at small $x_1$ 
({\it i.e.} large $x_2$) where no shadowing is possible. The thick solid curve 
in Fig.~\ref{eks} representing our result for the combined effect of
energy loss and shadowing confirms this conjecture. Indeed, it is below the
shadowing (dashed) curve at large $x_2$ exactly by the same amount as the
``shadowing'' curve from  Ref.~\cite{eks}.

The influence of energy loss on the effective nuclear suppression of
antiquarks found in \cite{eks} extends down to smaller $x_2$ (where most
of the E866 data are concentrated), giving to the $x_2$ distribution
quite a different shape compared to the shadowing we calculated. In
particular, nuclear suppression is strongest at the kinematical limit of
smallest $x_2$ in the E772/E866 experiments, while shadowing shown by
dashed curve vanishes in this limit.  We should again emphasize that
shadowing disappearing at small and large $x_2$ is a model-independent
expectation since it is based only on the uncertainty principle and
the kinematics of the DY reaction. 

Although the nuclear suppression from \cite{vasiliev} depicted in
Fig.~\ref{eks} by the thin solid line exceeds the calculated shadowing
effect shown by the dashed curve even at small $x_2\sim 0.015 - 0.02$
this is not the reason for our disagreement with Ref.~\cite{vasiliev}.
Indeed, as is stated above, we have checked that even with shadowing
twice as strong as the one shown by the dashed curve in Fig.~\ref{eks}
({\it i.e.} stronger at small $x_2$ than shadowing from Ref.~\cite{eks}),
the resulting energy loss is essentially the same.  We suppose that the
more important effect is the different shape suggested in \cite{eks} for
the $x_2$ dependence of the nuclear suppression, which does not vanish at
large $x_2$.

In conclusion, we believe that incorrect physical input at small $x_2$
and the particular antiquark density at large $x_2$ employed in the
analysis of Ref.~\cite{vasiliev} led to an incorrect result of vanishing
energy loss.  Although this result was confirmed with much worse
confidence by the recent analysis~\cite{nantes} of data from the NA3
experiment, it comes as no surprise that they arrived at the same
conclusions since their analysis was done essentially the same way as
that in \cite{vasiliev}.

\section{Summary and Conclusions }\label{conc}

We have presented a new analysis of the nuclear dependence of the E772/E866
Drell-Yan lepton pair production data.  This analysis makes use of a new
formulation of DY in the rest frame of the target, according to which the lepton
pair arises from the decay of a massive photon radiated by the incident quark in
a bremsstrahlung process.  We have been particularly interested in these data as
a source of information on the rate of energy loss of a quark propagating
through nuclear matter, encouraged by the observation that the rate of fall off
of the DY cross section data with $x_1$ near the point $x_1 =1$ is extremely
rapid and therefore sensitive to the amount of energy lost by the quark prior to
radiating the photon.  Since shadowing has a similar effect to that of energy
loss, namely suppression of the cross section at small $x_2$, it is important
that we have a good theoretical understanding of shadowing and that we
explicitly use this understanding to model the shadowing.

Identification of the dominant physical processes, as well as the
determination of their relative contribution, is guided by the notion of
coherence time, which can be interpreted as the lifetime of a virtual
fluctuation of the incident quark into the massive photon (and final
quark).  In particular, for short coherence times, the massive photon is
released immediately from the fluctuation as the incident quark scatters
from a bound nucleon. These processes clearly depend on the amount of
energy lost by the soft incident quark as it propagates from its point of
origin in a soft inelastic collision of the incident proton with a bound
nucleon to the point at which it undergoes bremsstrahlung, and these
processes are also sensitive to the energy lost by this quark for reasons
given above. Because the coherence time is short, the photon is radiated
immediately, and the final quark has no time to multiply scatter from other
nucleons before the radiation process has come to completion.  For long
coherence times, the dominant processes correspond to terms in which the
massive photon is radiated well in advance of any nuclear collision, and
shadowing becomes an issue as the final quark begins to scatter from other
nucleons of the nucleus. We describe shadowing quantitatively and without
free parameters in terms of a phenomenological color-dipole cross section
(fit to HERA data for the proton structure function), the nuclear thickness
function, and the longitudinal nuclear form factor.  Our model is shown to
reproduce shadowing where it has been observed, namely in DIS on nuclei
taken by the NMC collaboration.  The transition between the regime of long
and short coherence time is governed by the longitudinal form factor, with
the longitudinal momentum transfer bearing a simple inverse relationship to
the coherence length.
  
Our theory, as described above, contains one unknown parameter, namely the
rate of energy loss $\kappa$ of a quark propagating through nuclei.  Using
$\kappa$ as a free parameter, the theory was fit to the E772/E866 data for
DY lepton pair production arising from bombardment of $^2H$, Be, C, Ca, Fe,
and W targets by 800 GeV/c protons.  The data were binned both in $x_1$ and
$M$, a very important issue for reliable discrimination between the effects
of energy loss and shadowing. We found $\kappa = 2.28 \pm 0.31 GeV/fm$, in
close accord with our theoretical estimate of $2 GeV/fm$ determined by
considering string dynamics and gluon radiation.  Numerous tests confirming
the stability of our numerical analysis were made.  They also provided a
scale for the theoretical systematic uncertainty in the rate of energy
loss, which we estimated at about $\pm 0.5\,GeV/fm$.  Since the value we
found for $\kappa$ is substantially larger than the value reported in
previous work, we pointed out the important differences between the two 
analyses.

One should be cautious determining nuclear shadowing of sea quarks
from DY data which may be substantially contaminated by energy loss.
An essential demonstration our conclusions requires similar data for the DY
reaction on nuclei at lower energies, where shadowing is of no
importance but energy loss produces a stronger effect on the cross section. On
the other hand, at much higher energies of RHIC and LHC one can completely
disregard energy loss and test models for shadowing by direct comparison to
data. We expect spectacular shadowing effects producing a strong suppression of
the DY cross section for a wide range of $x_1$.

\medskip

\noindent {\bf Acknowledgment}: We are grateful to J\"org~H\"ufner,
Alberto~Polleri, J\"org~Raufeisen, Alexander Tarasov and Urs~Wiedemann
for numerous inspiring and clarifying discussions, and to Don~Geesaman
for valuable comments.  This work has been partially supported by a grant
from the Gesellschaft f\"ur Schwerionenforschung Darmstadt (GSI), grant
no. GSI-OR-SCH.

\end{document}